# Collaboration Planning of Stakeholders for Sustainable City Logistics Operations

**A Thesis**

**In**

**The Department**

**of**

**Concordia Institute for Information Systems Engineering (CIISE)**

Presented in Partial Fulfillment of the Requirements

for the Degree of Master of Applied Science (Quality Systems Engineering)

at Concordia University

Montreal, Quebec, Canada

April 2012

# CONCORDIA UNIVERSITY
## School of Graduate Studies

This is to certify that the thesis prepared

By:        Taiwo Olubunmi Adetiloye

Entitled:     **Collaboration planning of stakeholders for sustainable city logistics**

               **operations**

and submitted in partial fulfillment of the requirements for the degree of

**Master of Applied Science (Quality Systems Engineering)**

complies with the regulations of the University and meets the accepted standards with respect to originality and quality.

Signed by the final examining committee:

      \_\_\_\_\_\_\_\_\_\_\_\_\_\_\_\_\_\_\_\_\_\_\_\_\_\_\_\_ Chair
          Dr. Benjamin Fung

      \_\_\_\_\_\_\_\_\_\_\_\_\_\_\_\_\_\_\_\_\_\_\_\_\_\_\_\_ Internal Examiner
          Dr. Jamal Bentahar

      \_\_\_\_\_\_\_\_\_\_\_\_\_\_\_\_\_\_\_\_\_\_\_\_\_\_\_\_ External Examiner
          Dr. Navneet Vidyarthi

      \_\_\_\_\_\_\_\_\_\_\_\_\_\_\_\_\_\_\_\_\_\_\_\_\_\_\_\_ Supervisor
          Dr. Anjali Awasthi

Approved by:   \_\_\_\_\_\_\_\_\_\_\_\_\_\_\_\_\_\_\_\_\_\_\_\_\_\_\_\_\_\_\_\_\_\_
              Chair of Department or Graduate Program Director

\_\_\_\_\_\_\_\_\_\_\_\_\_\_\_2012     \_\_\_\_\_\_\_\_\_\_\_\_\_\_\_\_\_\_\_\_\_\_\_\_\_\_\_\_
                        Dean of Faculty

# Abstract

Collaboration planning of stakeholders for sustainable city logistics operations

Taiwo Olubunmi Adetiloye

**Concordia University**

City logistics involves movements of goods in urban areas respecting the municipal and administrative guidelines. The importance of city logistics is growing over the years especially with its role in minimizing traffic congestion and freeing up of public space for city residents. Collaboration is key to managing city logistics operations efficiently. Collaboration can take place in the form of goods consolidation, sharing of resources, information sharing, etc.

In this thesis, we investigate the problems of collaboration planning of stakeholders to achieve sustainable city logistics operations. Two categories of models are proposed to evaluate the collaboration strategies. At the macro level, we have the simplified collaboration square model and advance collaboration square model and at the micro level we have the operational level model. These collaboration decision making models, with their mathematical elaborations on business-to-business, business-to-customer, customer-to-business, and customer-to-customer provide roadmaps for evaluating the collaboration strategies of stakeholders for achieving sustainable city logistics operations attainable under non-chaotic situation and presumptions of human levity tendency.

City logistics stakeholders can strive to achieve effective collaboration strategies for sustainable city logistics operations by mitigating the uncertainty effect and understanding the theories behind the moving nature of the individual complexities of a city. To investigate system complexity, we propose axioms of uncertainty and use spider networks and system dynamics modeling to investigate system elements and their behavior over time.

The strength of the proposed work is its novelty and ability to investigate collaboration strategies both from macro- and micro-perspective allowing the decision maker to have a complete picture of the different possible collaboration opportunities and associated strategies and select the one most suited to their needs for sustainable operations planning.

# Acknowledgments

I wish to acknowledge my supervisor, Dr. Anjali Awasthi, for her guidance, patience and encouragements during the duration of my work. I also acknowledge the support of Concordia Institute for Information Systems Engineering (CIISE), Concordia University and its staff, in particular for the award of research assistantship.

I acknowledge my family for their supports and encouragements. Also, my fellow research students for being such wonderful companions and others, too numerous to mention, that made my studies at Concordia University a worthwhile and memorable experience.

My acknowledgments will be incomplete without reverence to the Creator the giver of my life, wisdom and strength and to his creatures who strive to overcome different hurdles of life and those helping to make the world a better and peaceful place to live.

I dedicate this work to

my parents, Catherine and Philip,

and to my elder brother, Charles.

# Table of Contents

# List of Figures

# List of Tables

# List of Acronyms

| | |
|---|---|
| ACS model | Advance collaboration Square model |
| CL | City logistics |
| FC | Full collaboration |
| HLT | Human Levity tendencies |
| NC | No collaboration |
| PC | Partial collaboration |
| SCS model | Simplified collaboration square model |
| SDM | System dynamics model/modeling |
| SN | Social network |
| UCE | Uncertainty effect |

# Chapter 1:

# Introduction

## 1.1. Background

City logistics(CL), as defined by Taniguchi *et al.* (1999, 2001), is "the process for totally optimizing the logistics and transport activities by private companies in urban areas while considering the traffic environment, the traffic congestion and energy consumption within the framework of a market economy". The main purpose of CL is to reduce city traffic congestion caused by freight-vehicle movement, improving vehicle utilization, and reducing emissions and pollutions without penalizing the city social and economic activities (Crainic *et al.*, 2011). Stathopoulos *et al.* (2011) emphasize that inefficient freight movements also contribute to noise, and increases in logistics that create hikes in product prices; also, stating that CL have a vital role to play in minimizing these negative impacts and ensuring freight movements within urban areas. The freight transport (Lorries > 3.5 tons) constitutes about 10% of total traffic within urban areas (Crainic and Sgalambro, 2009). Awasthi and Proth (2006) posit that the percentage will increase if the counts of delivery vans and cars are added. A city with high traffic of freight-vehicle movement, emissions and pollutions from moving freight-vehicles with the resultant effect on socio-economic activities of the city creates major obstacles to achieving sustainable CL operations.

It can be stated that the CL system complexity will either increase or decrease due to redundancy or emergence of collaborative subsystem(s) with respect to the change in the complexity of a city. Mathematically, this can be expressed as:

$$S_C = f(C) \tag{1}$$

Where $C = P_o \cup S \cup I_t \cup I \cup R \cup F_e \cup G \cup E \cup ...$

$$: \Delta S_C \propto \Delta C \tag{2}$$

$$\propto \Delta(P_o \cup S \cup I_t \cup I \cup R \cup F_e \cup G \cup E \cup ...)$$

$$= k_o \ (\Delta P_o \cup \Delta S \cup \Delta I_t \cup \Delta I \cup \Delta R \cup \Delta F_e \cup \Delta G \cup \Delta E \cup ...) \text{ - Associative property} \tag{3}$$

$$k_o \neq 1 \wedge 0: k_o \ \epsilon R$$

In this case, $S_C$ represents the system complexity defined as a function of the complexity($C$) of the city, where $C$ is an agglomeration or union of the individual complexities of the administrative policies ($P_o$), shippers' activities ($S$), information technology ($I_t$), infrastructures ($I$), residents' socio-cultural characteristics with demands ($R$), freights ($F_e$), goods ($G$), and environment ($E$) and so on[1] of a city while $k_o$ is the uncertainty variable with the assumption that its neutrality –"1"– and its nullity – "0" – hardly exist.

Hence, it can be inferred that the change in the subsystem complexity of a city is widely determined by variable factors which include the individual and interrelated complexities of the administrative policies, the activities of shippers, the information technology, the infrastructures in place, the residents' socio-cultural characteristics with demands, the freights, the goods, and the environment that change with time; which are under the subjective influence of uncertainty effect. The uncertainty effect can be described as unpredictability that often arises, the nature of which can be uncontrollable. For instance, such effect could be from the daily normal weather forecast of mild rain, snow, or dry weather to the extreme effect of man-made and natural disasters or even such things as an alien invasion, or a giant asteroid – that is, a planetoid – hitting a city. Also, with the exception of the change in information technology, which is more likely to be on the increasing trend, the rest of these variable factors can either be on the increase, decrease or in undulation with respect to time. This explanation corroborate the fact by Ovalle and Márquez (2003) that while internet e-collaboration tools play a vital role of "value creation enabler", they could at same time be signs of business and market "complexities" that companies will have to confront; and, in addition, they could also result in "intricate complexities" that would require in-depth analysis to assess their impacts on subsystems toward achieving sustainable CL operations.

---

[1] The major individual dynamic complexities are listed with an assumption that others may emerge in future.

## 1.2. Problem definition

The aim of this thesis is to provide a solution approach for collaboration planning of stakeholders for sustainable city logistics operations. To achieve this goal, we will address the following problems in our thesis:

1.     Understanding complexity of city logistics systems.

2.     Conceptualizing city logistics operations

3.    Investigating the role of Collaboration as enabler for sustainable city logistics

4.    Modeling and evaluating stakeholder collaboration strategies for sustainable city logistics planning

The first problem involves understanding the complexity of city logistics systems. We shall provide two mathematical models: one, to explain the uncertainty effect using axioms; and second, to categorize elements of city logistics system using spider networks.

The second problem involves conceptualizing city logistics operations. We propose to use system dynamics modeling to visualize the relationship between the elements of city logistics system; present reasonable explanations of how the dynamic nature of these elements contribute to pollution and congestions of city under the subjective influence of mild and extreme effect of uncertainty.

Thirdly, we will investigate the role of collaboration as enabler for sustainable city logistics. We will focus on four basic and key subsystems namely B2B, B2C, C2B, and C2C. Our approach require developing strategies for collaboration based on these key subsystems with two collaboration models known as simplified collaboration square model and advance collaboration square model. Also, we present some presumptions known as the human levity tendency and the trio-conditionality which are the non-chaotic, near chaotic and chaotic situations.

Finally, we will propose models for evaluating stakeholder collaboration strategies for sustainable city logistics. In one technique, we will assess socio-cultural characteristics(S), economy (E), and environmental (En) impacts at the macro-level of city logistics system and, in the other, heuristics based solution approach for decision making at the micro-level of city logistics system.

## 1.3. Thesis Outline

The rest of the thesis is organized as follows.

In chapter 2, we present the literature review on CL planning and collaboration strategies.

In chapter 3, we present our solution approach for collaboration planning of stakeholders for sustainable city logistics operations

In chapter 4, we present numerical analysis based on our solutions approach.

In chapter 5, we present the conclusions and future work.

# Chapter 2:

# Literature review

## 2.1. Sustainable city logistics systems

Taniguchi *et al.* (2003) state that global competiveness, efficiency, environment friendliness, congestion alleviation, security, safety, energy conservation, and labour force are the braces or cross bars that provide the goals or directions for city logistics to address urban freight transport issues, the essence of which must be considered to achieve the three pillars of the visions for city logistics which are sustainability, mobility and livability within urban areas. According to Crainic *et al.* (2009) "City logistics initiatives usually aim at reducing nuisances caused by freight transports in city while supporting social and economic development". Awasthi and Proth (2006) describe sustainable city logistics systems "as improving goods transport in urban areas through consolidation and coordination of goods transport activities to reduce the negative impacts of freight transport on city residents and their environment". This implies that sustainable CL systems can be achieved by collaboration of stakeholders involve in city operations. Langley (2000) posits that with the complexity and dynamic nature of today's rapidly evolving business world, any firm stands to lose if trying to "go it alone."

This makes it imperative to define sustainable city logistics as efforts aimed at improving goods transports in city through collaboration with regards to the uncertainty associated with the nature of the dynamic complexity of a city and the activities of stakeholders.

## 2.2. Stakeholders in city logistics systems

City logistics has many subsystems with key stakeholders or actors who comprise of shippers, residents, freight carriers, and administrator (Taniguchi *et al*., 2001) within an environment (see Table 1). Figure 2 describes the interactions between the key stakeholders involved in CL operations. The subsystems, connected through flows, are the different types of collaborative-commerce(c-commerce) that can be used in CL planning.  Gartner (1999), described c-commerce as "… dynamic collaboration among employees, business partners and customers throughout a trading community or market…" Consequently, the robust technologies of today provide for adaptation of e-commerce transactions through collaborations for CL planning and supply chain (SC).

**Table 1: City logistics stakeholders (adapted from Taniguchi *et al*., 2001).**

| CL Stakeholders | Description |
| --- | --- |
| Administrators | The Administrators represent the government or transport authorities' at the national, state or city level and whose objective is to resolve conflict between City Logistics authors, while facilitating sustainable development of urban areas. |
| Shippers | Comprise of manufacturers, wholesalers and retailers. |
| Residents | Are the consumers. |
| (Freight) Carriers | Comprise of transporters, warehouse, and companies. |

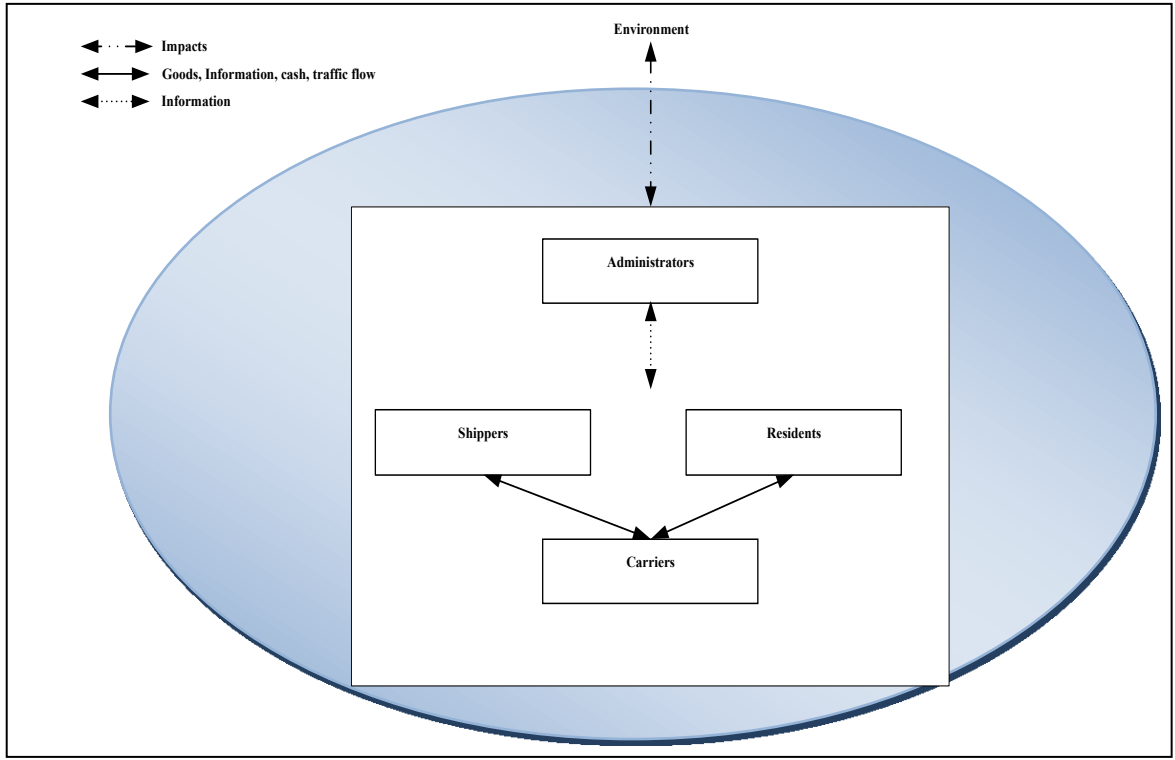

**Fig. 1: City Logistics stakeholders (source: ref. Fig. 1 –Awasthi and Proth, 2006)**

## 2.3 Decisions involved in city logistics planning

The decisions involved in CL planning include demand planning, vehicle routing and scheduling, impact assessment, fleet management, collaboration planning etc. These are discussed in details as follows:

### 2.3.1. Demand planning

Demand planning can be described as a business planning process that enables suppliers or shippers and retailers to make forecast on how much products need to be available to meet the demand of customers, clients or residents. Efforts aimed at having good demand plan can help reduce product wastage in inventory while also ensuring timely goods delivery to customers. Krajewski *et al.* (2007) state that leveraging points of sales data from the retailers can be done effectively provided the point of sales data at the goods companies is clean, integrated, synchronized, and harmonized with the internal data and other data by the retailer. The relevance of demand planning to collaborative planning, forecasting and replenishment (CPFR) was presented by Min and Yu (2007) from research on the value of information sharing among supply chain partners in relation to joint forecasting customer demand and co-managing business function.

Also, the influence of demand uncertainty cannot be underemphasized when looking at how products demand fluctuates from time to time (Sultana and Shathi, 2010). Zhao *et al.* (2002a, 2002b) presents findings that can help supply chain managers forecast supply chain performance with development of a simulation model that examines demand forecasting and inventory replenishment decisions by retailers and product decisions by the supplier under demand uncertainty.  In relation to city logistics planning, Crainic *et al.* (2011) emphasize the need to reduce congestion and environmental impacts caused by freight-vehicle movements, without infringing on the social and economic activities in the city; a premise that necessitated developing a model for demand uncertainty in two-tiered city logistics planning.

### 2.3.2. Vehicle routing and scheduling

Vehicle routing and scheduling was proposed by Dantzig and Ramser (1959) in the truck dispatch problem. It is a combinatorial optimization problem. To address this problem, the used methods involve integer programing that seeks the global minimum route for delivery of goods vehicles to customers while implicitly minimizing cost of distributing the goods.

In city logistics operations, vehicle routing and scheduling has become an important technique. It is used not only for cost optimization but also addressing the problems of congestion and environmental pollution within the city. Taniguchi *et al.* (2001) recommend vehicle routing and scheduling models as a core technique for addressing city logistics problems. A variance of the vehicle routing and scheduling respecting the constraints of delivery time windows (VRPTW) and government regulations was used by Awasthi and Proth (2006). They present a system based approach for city logistics decision making with the layout of a simulation model called CILOSIM. Barceló *et al.* (2005) report a modeling framework supported by computer decision support system taking into account the nature of problems addressed in city logistics with a graphical user interface that automatically generates network location and vehicle routes. Galic *et al.* (2005) conceived the programming language MARS with routing oriented and built-in data types and instructions to enable fast development and testing of constructive and heuristic algorithms, distributed and parallel execution of the algorithms in the cluster/grid environment and assistance in practical VRP problem solving. Xu *et al.* (2011) present vehicle routing optimization with soft time windows in a fuzzy random environment (VRPSTW) with two objectives of minimizing the total travel cost and maximizing the average satisfaction level of all customers.

### 2.3.3. Fleet management

The management of fleets (vehicles) is known as Fleet Management. For examples, the management of commercial motor vehicles, cars and trucks. Fleet management comprises of the following range of functions:

- Vehicle financing

- Vehicle maintenance

- Vehicle telematics (tracking and diagnosis)

- Driver management

- Speed management

- Fuel management

- Health and safety management

The objective of fleet management includes:

- Reducing or minimizing risks associated with vehicle investments for companies which depend on fleets for their business.

- Ensuring there are significant reductions in overall transportation and staff costs.

- It allows companies to improve the efficient and productivity of staffs and their vehicles.

- It ensures full compliances with government legislations.

*(Wikipedia, Fleet management)*

Fleet management has been greatly enhanced by recent advances in Information and Communication Technology (ICT) as well as Intelligent Transport Systems (ITS) (Taniguchi *et al.*, 2005). For example, those used for vehicle tracking with GPS, diagnostics of vehicle mechanical fault, monitoring driver's behavior, remotely disabling a vehicle, and fleet management software etc.

Several researches have been directed towards fleet management. Crainic and Laporte (1998) present theoretical bases on fleet management and logistics. The development and evaluation of an intelligent transport system for city logistics was reported by Zeimpeki *et al.*(2008) detailing how the system handles unforeseen events resulting from vehicle delivery executions in real time in the dynamics of a city logistics environment. Awasthi *et al.*(2011) present a centralized fleet management system(CFMS) for cybernetic vehicles called cybercars providing approach that confronts the challenges before CFMS which include conflict-free routing, accommodating immediate request from customers, empty cybercars to new services or parking stations and those running below their threshold battery levels to recharging stations. Wang (1991) present a fleet management system to be used in freight forwarding business, where cargos trucks are to be monitored and dispatched in real time manner via a GIS/GPS platform.

### 2.3.4. Impact assessment

Impact assessment provides performance indicators that can help city logistics planners in addressing the problems of congestions and freight-vehicle emissions-causing-pollutions in the city. Taniguchi *et al.* (2001) classified impacts into economy, energy, environmental, financial, and social with models for assessing their effects on the city.

Environmental pollution, emission, CO, NO$_x$ etc. from freight transports, the impact of which poses great concerns to decision makers at city administrative level (Gragnani *et al.*, 2003; Taylor; and, Awasthi and Proth, 2006), are often the direct and indirect consequences of other identified impacts resulting from the activities of city-dwellers. Marquez *et al.* (2003) present a policy oriented model for the impact assessment of urban goods movement in relation to energy consumption, congestion and environmental concerns (pollution, greenhouse gas and noise) to help in local decision making. An environmental impact assessment model of urban goods movement aimed at local decision makers with plans for achieving sustainable city development was developed by Segalou *et al.*, 2003.

### 2.3.5. Collaboration planning

Barratt (2004) investigated crucial questions on collaboration such as: Why do we need to collaborate? Where can we collaborate and with whom should we collaborate? Over what activities can we collaborate? The peculiarity of answers provided by academicians and practitioners is often on the premise that two or more independent firms can achieve greater success if they synergize their supply chain processes with the goal of creating value to end customers and stakeholders than acting alone (Horvath, 2001; Simatupang and Sridharan, 2002; and Simatupang *et al*, 2004).

While, collaboration seems to be the consensus agreed upon by authors for achieving sustainable city logistics systems; uncertainty hinders the prospect of achieving a perfect city logistics. Figure 2 illustrates the Venn diagram of stakeholders, city logistics and uncertainty.

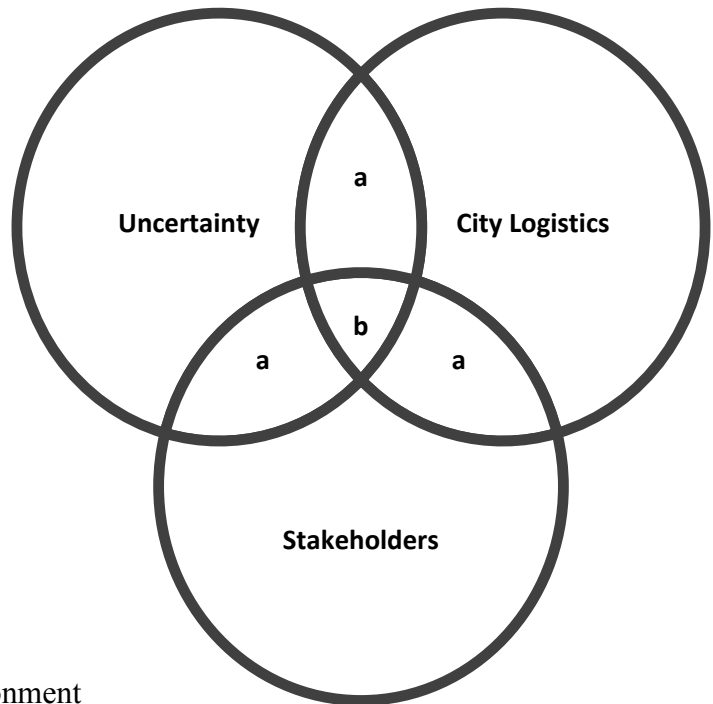

a- Environment

b- City complexity

**Fig. 2: Venn diagram of stakeholders, city logistics and uncertainty**

Therefore, it can be stated that collaboration of stakeholders in an environment within a city are predicated on the subjective influence of uncertainty effect.

Cassivi (2006) present analysis of e-collaboration tools with regards to different partners along the supply chain; and made a categorization of firms according to their level of collaboration within a supply chain environment. Holweg *et al.* (2005) research on the classification of collaboration initiatives using conceptual water-tank approach; and discuss their dynamic behaviors and key characteristics. Barratt (2004) address the difficulty expressed by some authors on implementing supply chain collaboration by proposing approach to segment supply chain based on customer buying behavior and service needs; and to identify the elements that make up supply chain as well as the interrelationship among the cultural, strategic and implementation elements of supply chain.

## 2.4. Types of collaboration

Collaboration, in supply chain, is often categorized in terms of scope and e-commerce.

In terms of scope, Simatupang and Sridharan (2002) divided supply chain collaboration into two categories: the first which is vertical could include collaboration with customers, internally (across functions) and with suppliers; and the second, which is horizontal could include collaboration with competitors, internally and with non- competitors. In terms of ecommerce, there could be:

### *Business-to-business (B2B)*

Business-to-business (B2B) describes transactions between commercial entities, such as between a manufacturer and a wholesaler, or between a wholesaler and a retailer. It usually involves purchase of millions of components or products by an organization from multiple sources of supply (Plant, 2000).

### *Business-to-consumer (B2C)*

Business-to-consumer (B2C), also called Business-to- Customer, can be considered as that by which organizations transact business with customers through retail portals. Common examples of B2C-based organizations are Amazon.com, Wal-Mart and eBay.

### *Business-to-employee (B2E)*

Business-to-employee (B2E) uses an intra-business network that links businesses to employees to facilitate transactions of products and services. Such transactions are rendered with automate employee-related corporate processes. Examples of B2E applications include

- Online insurance policy management
- Corporate announcement dissemination
- Online supply requests

(Wikipedia, Business-to-employee)

### *Business-to-government (B2G)*

Business-to-government (B2G) referred to as a market definition of "public sector marketing" which encompasses enterprise marketing products and services to government establishments at the federal, state and local level through integrated marketing communications techniques such as strategic public relations, branding, advertising, and web-based communications (Plant, 2000).

### Business-to-manager (B2M)

Business-to-manager (B2M) enables transactions between enterprises that have products and services and professional managers. Information gathering on the net with B2M scheme are usually done for earning commission by providing services for enterprises. (Wikipedia, Business-to-manager)

### Customer-to-business (C2B)

Consumer-to-business (C2B) is an inversion of the B2C. The C2B is an electronically facilitated transaction for customers to provide a new idea that can be used by businesses for new product development. Examples of this can be an inventor that puts his/her invention for sale and auction to a firm for production and sales via net portal.

### Customer-to-customer (C2C)

Consumer-to-consumer (C2C) (or *citizen-to-citizen*) is an electronically facilitated transaction between consumers that eliminate the need for middle men. A Customer directly contacts another customer regarding his/her products or services. An example is apartment rental on craigslist.com.

### Government-to-Business (G2B)

Government-to-Business (abbreviated G2B) can be considered as a type of ecommerce by which Governments interacts with businesses (or citizens) through government portal (Plant, 2000).

For fuller discussions on different types of ecommerce, readers should refer to Nemat (2011).

## 2.5. Methodologies for city logistics planning

The methodologies used by various authors in dealing with the problems of CL systems can be broadly classified into two main categories: qualitative and quantitative (see Figure 3).

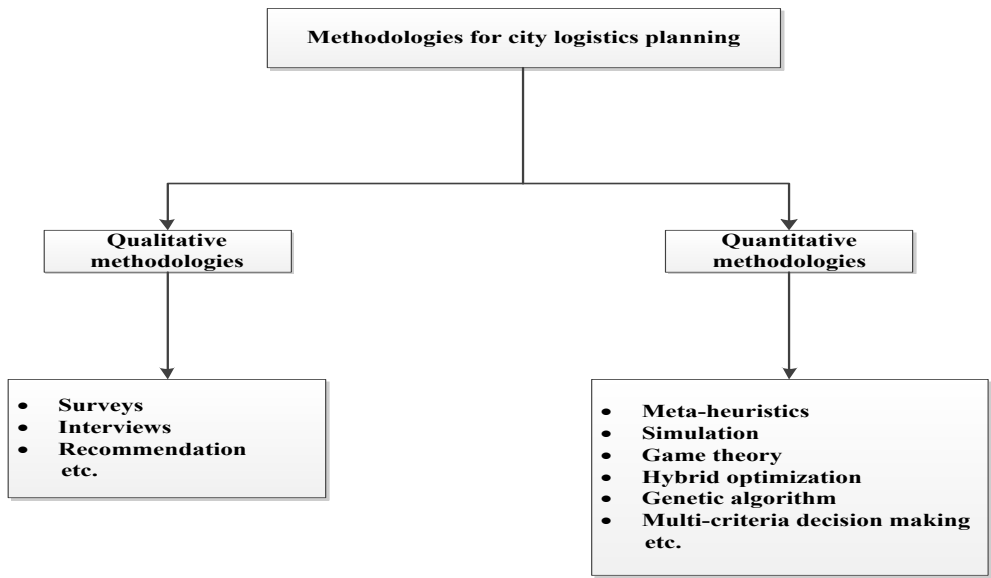

**Fig. 3: Methodologies for city logistics planning**

These categories are described in detail as follows:

### 2.5.1. Qualitative

The qualitative methods primarily involve the use of surveys, interviews, and recommendation in addressing the issues of CL systems. They are usually inquiries that seek to find descriptions or distinctions based on some characteristics rather than on some quantity or measured value.

### 2.5.1.1. Surveys

Survey methodology for CL systems studies the sampling of shippers and freight carriers as well as goods transports as constituents of CL systems with a view of making statistical inferences about the elements using the sample. Muñuzuri *et al.* (2005) present a compilation of initiatives that can be implemented by local administrations in order to improve freight deliveries in urban environments.

### 2.5.1.2. Interview

Interview is used for gathering vital information on CL systems with a view to knowing the general opinions of city dwellers on what improvements can be made within the city environment. Interview can also help in knowing the opinions of stakeholders over the questions of collaboration which according to Barratt (2004) are: Why do we need to collaborate? Where can we collaborate and with whom should we collaborate? Over what activities can we collaborate? The answers gotten from interviews can form an opinion poll useful for making statistical inferences.

### 2.5.1.3. Recommendation

In city logistics operations, there is often the need to have a worth of confidence or acceptance among stakeholders. This requires providing recommendation for an individual to an organization usually based on his or her ability to exercise competency and trust in the establishment. William (2004) presents an instance of a recommendation policy document included in the regional transportation plan for the San Francisco area, to be adopted in 2005.

### 2.5.2. Quantitative

The quantitative methodologies rely on numeric or mathematical techniques for problem solving. For example, meta-heuristics, game theory, simulation etc.

### 2.5.2.1. Meta-heuristics

Meta-heuristics is a high–level general strategy which guides other computational heuristics methods to find optimal solution to a combinatorial problem by searching iteratively over a discrete space or trying to improve a candidate solution with regard to a given measure of quality. Examples of meta-heuristics are travelling salesman problem, tabu search, simulated annealing, genetic algorithms and memetic algorithms.

Meta-heuristics are used for finding an optimal solution in combinatorial problems over a discrete space, such as in logistics network. Cordeau and Maischberger (2009) present a parallel iterated tabu search heuristics for solving four different vehicle routing problems. Crainic *et al.* (2011) present a very efficient method for solving a series of stochastic capacitated multi-commodity network design (CMND) problems when compared to direct solution approach using latex version of CPLEX with progressive hedging-based meta-heuristics for stochastic network design. Pedersen *et al.* (2009) present models and tabu search meta-heuristics for service network design with asset-balance requirements; offering both arc-and cycle-based formulations for the models and tabu search meta-heuristics framework for the arc-based formulation. Liu *et al.* (2010) present memetic algorithms for the solution of the task selection and routing problem with full truckload that addresses the task selection and routing problems in collaborative truckload transportation.

### 2.5.2.2. Game theory

Game theory is formally defined as "the study of mathematical models of conflict and cooperation between intelligent rational decision makers" (Myerson, 1991). It is method for studying the strategic decision making required for collaboration among decision makers. The types of games include:

1. Combinatorial games

2. Cooperative and non-cooperative games

3. Discrete and continuous games

4. Infinite long games

5. Many-player and population games

6. Metagames

7. Simultaneous and sequential

8. Stochastic outcomes( and relation to other fields)

9. Symmetric and asymmetric

10. Perfect information and imperfect information

11. Zero-sum and non-zero sum

Game Theory is mainly a logic and/or optimization mathematical method for analysis of human behavior; for example, in analyzing friends or foes behavioral characteristics during a collaborative experiment (List, 2006). Game theory has been used in different areas of science, making it one of the most widely researched topics for understanding human behaviors. In game theory, there is the Nash equilibrium (named after John Forbes Nash, who proposed it) a solution concept of a game involving two or more players, in which each player is assumed to know the equilibrium strategies of the other players, and no player has to change his own strategy unilaterally since such results in no gain for the player that deviates (Osborne and Rubinstein, 1994).

Mathematically, the Nash equilibrium states that let *(S, f)* be a game with *n* players, where $S_i$ is the strategy set for player *i, $S = S_1 \times S_2 ... \times S_n$* is the set of strategy profiles and *f = ($f_1(x)$... $f_n$ (x))* is the payoff function for *x $\in$ S*, then a strategy profile *$x^*$ $\in$ S* is a Nash equilibrium (*NE*) if

$$\forall_i, x_i \in S_i, x_i \neq x_i^* : f_i(x_i^*, x_{-i}^*) \geq f_i(x_i, x_{-i}^*)$$

Where $x_i$ is taken to be a strategy profile of player *i* and $x_{-i}$ is the strategy profile of all players except for player *i*. Examples of applications of the Nash equilibrium can be found in the prisoner dilemma, coordination game, network traffic, and competition game.

Bell (2000) presents a game theory approach to measure the performance reliability of transport network; upon which the Nash equilibrium measures network performance providing rationality for user to make cautious evaluation of network design. Also, Bell and Cassir (2002) introduce an application of the game theory using the mixed-strategy Nash equilibrium to describe a risk-averse user equilibrium traffic assignment. Uchiyama and Taniguchi (2010) present a route choice model based on evolutionary game theory considering the travel time reliability and traffic impediment; a model that provides a basis for identifying the route a dispatcher or shipper chooses in consideration of the change of the daily route times and the situation of the traffic impediments using measured data. A game theoretic conceptual framework model was presented by Roumboutsos and Kapros (2008) to highlight strategies undertaken by public transport operators, public or private, vis-à-vis operational integration strategies with the Nash equilibrium used to identify possible outcomes in various situation.

### 2.5.2.3. Genetic algorithm

Genetic algorithm is a search heuristics that is routinely used to generate useful solutions for optimization and search problems using a method that mimics natural evolution, namely: inheritance, selection, mutation and crossover. Using genetic algorithm in solving optimization and search problems, a better solution approach can be achieved when compared to other approaches primarily using dynamic programming which are found to be computationally intensive as maintenance infrastructure elements increase in urban areas. The main disadvantage of genetic algorithm is that as the complexity of the problem increase, performance of the genetic algorithm tends to become NP-hard (Jha *et al.*, 2005).

Jha *et al.* (2005) presents a genetic algorithm-based decision support system for transportation infrastructure management in urban areas as a solution approach for two models developed for, one, minimizing inspection travel time and the other for obtaining an optimal maintenance schedule over a planning horizon.  Yang *et al.* (2005) present location model and genetic algorithms for optimizing the size and special distribution of city logistics terminals with the goal of minimizing the total freight transport cost in the city.

**2.5.2.4. Hybrid optimization**

Hybrid optimization essentially uses two or more optimization algorithms to solve same optimization problem. It involves the combination of different optimization algorithms which include:

1.  Combinatorial programming e.g. greedy and hungarian algorithms

2.  Evolution computation e.g. genetic algorithm, memetic algorithm, swarm algorithm

3.  Linear programming

4.  Nonlinear optimization

5.  Newton method in optimization

6.  Dynamic programming

7.  Nearest neighbor search

8.  Simulated annealing

9.  Stochastic tunneling

10. Local search

For a full list of the optimization algorithms the reader should, please, refer to Wikipedia, list of algorithms (in references).

In application of hybrid optimization to city logistics systems, Omrani *et al.* (2009) develop a hybrid approach for evaluating environmental impacts for urban transportation mode sharing. Chen *et al.* (2010) present a routing optimization algorithm in city logistics distribution by adopting a framework of genetic algorithm with a strong local search ability greedy algorithm for providing a mix of genetic crossover operator and greedy crossover operator to achieve rapid convergence effects that improves the performance of the local search genetic algorithm. Simão et al (2009) apply dynamic programming algorithm, merging math programming with machine learning, for large scale fleet management to present a solution with extremely high-dimensional state variables.

### 2.5.2.5. Simulation

According to Rossetti (2010), the main purpose of a simulation is to allow observations about a particular system as a function of time with the key advantage of modeling entire systems and complex relationships; which, makes it possible to model Real-world systems that are often too complex. For example, a Real-world system within a B2B subsystem can be a distribution network of plants, transportations links and warehouses involving stakeholders at same or different areas of the system (CL) operations within an environment. However, the reality of a complete or perfect simulation model of Real-world system can be unattainable due to the moving nature of the complexity and the uncertainty effect.

In CL system research, Russo and Cartenì (2005) present a tour-based approach that reproduces the choice structure of freight transport; an approach that simulates the dependences existing between successive trips of the same distribution channel. Barceló *et al.*(2005) present a methodological proposal based on an integration of vehicle routing and dynamic traffic simulation models that emulate the actual traffic conditions to provide optimal dynamic routing and scheduling for vehicle under test in the European Project MEROPE of the INTERREG IIIB Programme, and in the national project SADERYL. A case study was presented by Taylor for Sydney involving the evaluation of likely impacts of transport policies aimed at mitigating the environmental impacts of urban freight transport using city logistic systems simulation model that optimizes logistics and efficiency under congested urban traffic conditions.

### 2.5.2.6. Multi-criteria decision making

Multi-criteria decision making is a sub-discipline of operations research that is concerned with structuring and solving problems with multi-criteria instance. It was introduced in the early 1960's and has since attracted a number of contributions to theories and models. Research into multi-criteria decision making has involved the addition of Fuzzy set allowing for a better solution approach for problems that have hitherto been inaccessible and unsolvable with standard multi-criteria decision making techniques (Carlsson and Fullér, 1996). Fuzzy theory was introduced by Zadeh (1965) to deal with uncertainty and ambiguity of data; for instance, as might be related to decision making. Improvement in multi-criteria decision making techniques have been gotten from knowledge in many fields such as economics, software engineering, mathematics, behavioral decision theory, and so on( Wikipedia, Multi-criteria decision analysis).

Location planning for urban distribution centers under uncertainty has been presented using multi-criteria decision making approach (Awasthi *et al.*, 2011). Multi-criteria models and approach that address location problems have been researched by Lee *et al.* (1981); Ross and Soland (1980); Puerto and Fernandez (1999); and Erkut *et al.* (2008). Fuzzy set multi-criteria decision making models for location problems have been researched by Anagnostopoulos *et al.* (2008); Ishii *et al.* (2007); and Kahraman *et al.* (2003).

## 2.6. Summary

Table 2, provides a summary of some research works by various authors on CL operations.

**Table 2: Summary of some research on CL planning**

| Author(s) | Problem addressed | Problem type | Difficulty level |
|---|---|---|---|
| Uchiyama and Taniguchi, 2010. | A route choice model based on evolutionary game theory considering the travel time reliability and traffic impediment. | Stochastic | Moderate |
| Awasthi and Proth, 2006 | A systems-based approach for city logistics decision making. | Deterministic | Moderate to high |
| - Crainic et al, 2011.<br>- Crainic et al, 2009. | - Progressive hedging-based meta-heuristics for stochastic network design;<br>- Service network design models for two-tier city logistics. | Stochastic. | Moderate to high |
| - Goldratt, 2010.<br>- Marton and Paulová, 2010. | - Theory of constraints;<br>- Applying the theory of constraints in the course of process improvement. | Stochastic,<br>Deterministic | Moderate |
| Lin et al, 2010. | Task selection and routing problems in collaborative truckload transportation. | Heuristic,<br>Stochastic | Moderate to high |
| Kale et al, 2007. | Analyzing private communities on Internet-based collaborative transportation networks. | Deterministic | Moderate |
| Barceló et al, 2005 | Vehicle routing and scheduling models, simulation and city logistics. | Heuristic,<br>Stochastic | Moderate to high |
| Jha et al, 2005. | A genetic algorithms-based decision support system for transportation infrastructure management in urban areas. | Search heuristic,<br>Stochastic | Moderate to high |
| - Zhou et al, 2010.<br>- Taylor. | - Strategic alliance in freight consolidation<br>- The city logistics paradigm for urban freight transport. | Stochastic,<br>Deterministic | Moderate |
| Shang and Marlow, 2005. | Logistics capabilities and performance in Taiwan's major manufacturing firms. | Deterministic | Moderate to high |

**Table 2: Summary of some research on CL planning**

| Solution Approach | Key Advantage | Key Disadvantage |
|---|---|---|
| Game theory | Provides a strategy for analyses of human behaviors as a reflection of performance in the theories of game. | Mainly a logic and/or optimization mathematical method. |
| System dynamic model | It allows for visualization of the necessary components of CL operations. | Does not allow for the micro-evaluation of strategies. |
| Pseudo-Backhaul routing concept; Decomposition approach | Very efficient method for solving a series of stochastic CMND problems when compared to direct solution approach using latex version of CPLEX. | More work is needed for extending and refining the solution approach. |
| Theory of Constraints | Allows for fostering teamwork of stakeholders, making them aware of the constraints and the need for integration to assist the constraint process. | Can be difficult to apply if the constraint process is constantly moving. |
| Memetic algorithm (MA) for the solution of the task selection and routing problem with full truckload. | The proposed MA outperforms the classical genetic algorithm in terms of solution quality i.e. it can provide good solutions in reasonable computation time. | The task selection and routing problem with full truckload is NP-hard. |
| Transportation exchange model | A transportation model for analyzing cost and benefits of collaborative community to shippers and carriers which facilitates information exchange between carriers and shippers with private transportation community. | There has been mixed success over the new trends of private transportation community due to uncertainty of cost and benefits to shippers and carriers. |
| Vehicle routing and scheduling model. | A very good approach for finding optimal solution in logistics and transportations, e.g. objective of minimizing the cost of distributing goods. | Finding global minimum for these types of problems is computationally complex. |
| Genetic Algorithm | A better solution approach can be achieved when compared to other approaches primarily using dynamic programming which are found to be computationally intensive. | As the complexity of the problem increase, performance of the genetic algorithm has to be examined since the problem may become NP-hard. |
| Simulation Modeling | Allows modeling of entire systems and complex relationships as a function of time; which makes it possible to model Real-world systems that are often too complex. | There can be no perfect simulation model due to moving nature of the complexity and uncertainty inherent in most dynamic systems. |
| Structural Equation Modeling | Offers more meaningful and valid results in complex data analysis that can be better than alternative methods. | More effort is necessary until the greater complexity can be achieved. |

# Chapter 3:

# Solution approach

## 3.1. Understanding complexities of city logistic systems

Mathematical modeling is a useful way of describing a system using mathematical concepts, and expressions. Many scientists have based their explanations of collaboration by stakeholders involved in CL operations with the use of mathematical models. In this research work, two mathematical models are proposed for understanding the complexities of city logistic systems under the following topics:

1. Uncertainty effect( Axioms)

2. Categorizing city logistics elements( using Spider networks)

With these models, to be discussed in the following subsections, we shall provide a rational for CL systems, the constraints against such collaborative efforts and heuristics to mitigate the impact of these constraints in order to achieve sustainable CL operations.

### 3.1.1. Uncertainty effect

The uncertainty effect, in the context of CL and other deductions thereof, can be measured by the uncertainty variable, $k_o$. A mathematical diagnosis of $k_o$ revealed that its evaluation can be described by assigning binary values to the unpredictability effect of the conditions of air, wind, snow, and dry weather, as well as that of tornado, hurricane, planetoid (hitting a city), alien invasion and so on.

The validity of this argument is based on two basic axioms:

1. The neutrality "1" and nullity "0" of $k_o$ hardly exist.

   $$k_o \neq 1 \ \wedge \ 0: \ k_o \ \epsilon \ R$$

2. $k_o$ is an effector[2] ( or effect vector) such that

   $$: k_o = |\overrightarrow{\overrightarrow{k_o}}| = |\overrightarrow{\overrightarrow{k_1}}n_1 + \overrightarrow{\overrightarrow{k_2}}n_2 + \overrightarrow{\overrightarrow{k_3}}n_3 + \overrightarrow{\overrightarrow{k_4}}n_4 + \cdots|$$

That is, the effectors of $k_o$ is the summation of individual condition effectors, $k_1\,n_1, k_2n_2, k_3\,n_3, k_4\,n_4$ etc.

The constraints necessary for axiom 2 to satisfy axiom 1 are:

   a.      $k_1, k_2, k_3, k_4 \ldots \epsilon \ R \neq 1 \ \wedge \ 0$

   b.      $S$calars $n_1, n_2, n_3, n_4 \ldots \epsilon \ N$ ; cannot be equal to zero, simultaneously

An effector can either be negative or positive; while a condition can be defined as matters such as air, water, wind, nuclear energy etc. found within the earth sphere, and perceived matters such as sunlight, planetoid, aliens as well as matters in black hole, etc. that exist beyond our earth sphere, somewhere, in outer space. Conditions are identified as both negative and position effectors; while, conditions such as tornado, hurricane, planetoid, and black hole are primarily negative effectors. Hence, it can be stated that a negative effector causes a harmful, uncertainty effect while a positive effector causes an innocuous, uncertainty effect with impact on the collaboration of CL stakeholders within a city. For example, a mild wind may have a negative effect if there is air pollution and with multiplicity – that is, a strong wind – it can endanger a city, thereby militating against collaboration of stakeholders involved in CL operations. Positive effectors often do not endanger a city; rather, they facilitate collaboration of stakeholders involved in CL operations but can be inhibitive with very high multiplicity. The following Table 3 describes the "hypothetic values" assigned for some of the condition effectors.

---

[2] An effector or effect vector is different from a vector because a vector has magnitude and director but an effector has magnitude and (positive and negative) effect.

The hypothetic values assigned for the conditions are based on the following assumptions:

i.   The level of human perception of the severity of a condition. For instance, air is perceived as least severe –hence, it is assigned the smallest hypothetic value –while black hole is perceived as most severe with the highest hypothetic value.

ii.  Conditions deemed as opposite in nature, such as wet and dry weather, are assigned same hypothetic value.

iii. Conditions on the same scale of effect such as light and planetoid which emits photons and heat energies are assigned same binary numbers before the decimal point but are differentiated with different binaries numbers after the decimal point.

iv.  The further a condition exist from the earth sphere the greater the hypothetical value.

**Table 3: Hypothetic values for condition effectors**

| $x$ | Condition | Negative effector, $-k_x n_x$ | Positive effector, $k_x n_x$ |
|---|---|---|---|
| 1 | Air | $-1.1n_1$ | $1.1n_1$ |
| 2 | Dry | $-1.11n_2$ | $1.11n_2$ |
| 3 | Wet | $-1.11n_3$ | $1.11n_3$ |
| 4 | Wind | $-11.1n_4$ | $11.1n_4$ |
| 5 | Snow | $-111.1n_5$ | $111.1n_5$ |
| 6 | Water | $-111.11n_6$ | $111.11n_6$ |
| 7 | Tornado | $-1111.1n_7$ | $-$ |
| 8 | Hurricane | $-1111.11n_8$ | $-$ |
| . | . | . | . |
| . | . | . | . |
| . | . | . | . |
| $a-3$ | Planetoid | $-11111.11n_{a-3}$ | $-$ |
| $a-2$ | Sunlight | $-11111.1111n_{a-2}$ | $11111.1111n_{a-2}$ |
| $a-1$ | Alien invasion | $-11111111.1n_{a-1}$ | $11111111.1n_{a-1}$ |
| $a$ | Black hole | $-11111111111.1n_a$ | $-$ |

Also, the choice of binary numbers with "1s" as hypothetic values is from the fact that the arithmetic of binary number are better solved by computers than by the human mental cognition system due to the boredom and difficulty that become evident as its arithmetic complications unfolds. Secondly, binary numbers "1" and "0" digits are used to depict switch states and for the hypothetic values of condition effectors, though initialize with "1s" as a representation of a unique perfect state, a mixture of "1s" and "0s" often emerge as complex number in the axioms of the uncertainty affect. Moreover, attempting to solve an axiom of uncertainty problem with binary numbers tends toward an unpredictable complex numeric that best mimic the uncertainty effect.

By this discussion, a perception has been created on the impacts of uncertainty effect on the individual complexities of a city, and how it influences the collaboration of stakeholders involve in CL operations.

### 3.1.2. Categorizing city logistics elements

A spider's web is a spiraling polygon for a good reason, which is to meet the needs of the spider; albeit, the dexterity at which the spider holds together its web or network (at the center) and yet preys an insect is unpredictable or uncertain (please, see Figure 4). This statement provides an analogy for categorizing the elements of city logistics systems. The following Figure 5 provides a diagrammatic representation of equations (1-3), earlier discussed in the introductory section, to show the relationship between uncertainty effect and individual complexities of the city.

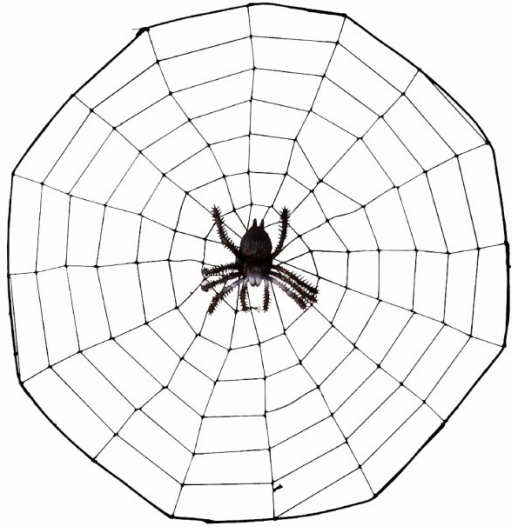

**Fig. 4: Spider web**

The figure could be seen to be similar to the outer polygon of the spider network. The individual complexities of a city ($\Delta C$) , that includes the change in the administrative and government policies ($\Delta P_o$), shippers'activities ($\Delta S$), information technology ($\Delta I_t$), infrastructures ($\Delta I$), residents' social and cultural values with demands ($\Delta R$), freights ($\Delta F_e$), goods ($\Delta G$), and environment ($\Delta E$) of a city, form the edges (or nodes) of the octagonal spider network. The bi-directional connections or links between these complexities is classified as tangible or intangible path or connector. Tangible links are directly measurable while intangible links are not measurable since they lack lucid physical appearance; however both connectors follows a clockwise, anticlockwise or dual-wise direction that signifies the agglomeration of the individual dynamic complexities of the city. Also, an edge having two tangible paths directly at opposite sides can be called tangible node, which is typically identified as $\Delta R$; likewise, an edge with two intangible paths directly at opposite sides can be called intangible nodes, consisting $\Delta P_o$, $\Delta I_t$ and $\Delta E$. while an edge with a tangible and intangible links directly on its opposite sides can be called a semi-tangible or semi-intangible node, which consist of $\Delta S$ , $\Delta G$, $\Delta I$, and $\Delta F_e$.

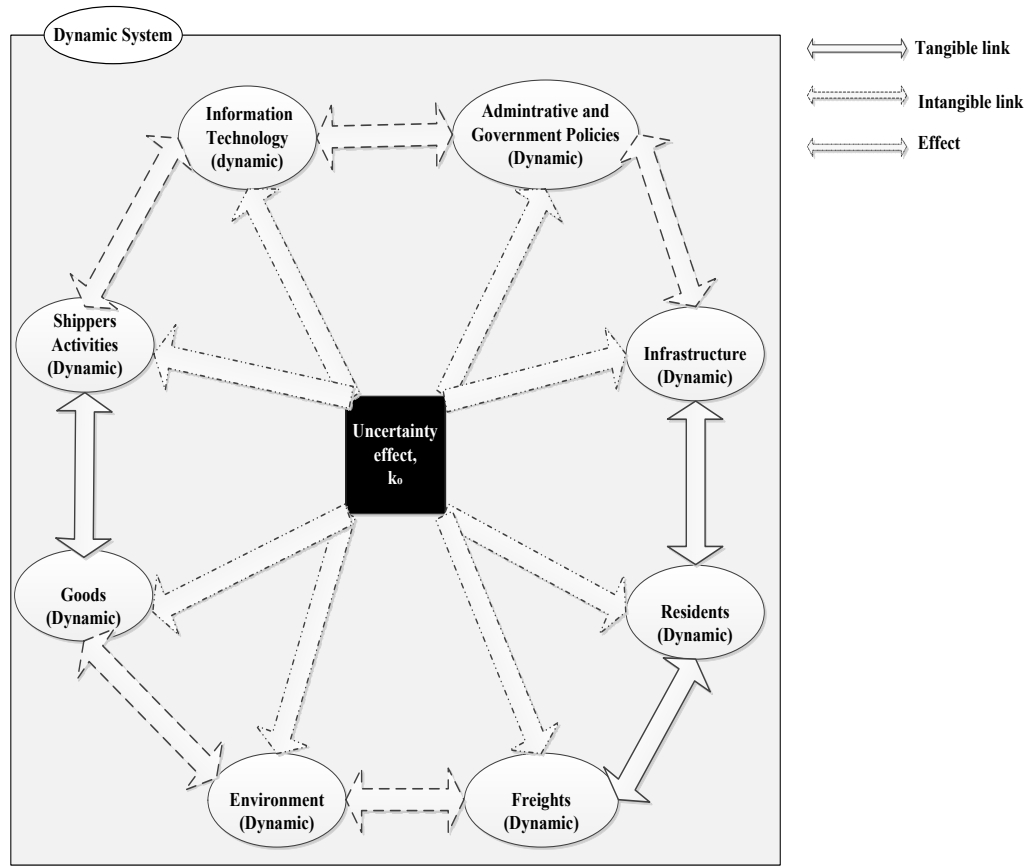

**Fig. 5: Spider network –linkage between uncertainty effect and individual complexities of a city**

Also, unidirectional links extend from the black-box at the center towards each of the dynamic complexities located at the edges of the octagon. A unidirectional link is considered to be the influence that uncertainty effect exerts on the individual complexities of the city. The uncertainty effect represented by $k_o$ is depicted by the black-box at the center; while the dynamic system complexity, $\Delta Sc$, is depicted by the larger grey box enclosing the whole spider network.

Table 4 further illustrates the components of the spider network.

**Table 4:  Components of the spider network.**

| Links | Nodes | Node instance | Examples |
|---|---|---|---|
| Tangible | Tangible | Residents | Customers |
| Intangible | Intangible | Administrative and Government policies | Tax, structuring, regulatory policies etc. |
| | | Environment | Lands, air, rivers, climate etc. |
| | | Information Technology | Social networking, e-collaboration etc. |
| Semi-tangible | Semi-tangible | Freights | Vehicles, ships, air cargo, railway. |
| | | Shippers | Manufacturer, wholesaler and retailers. |
| | | Goods | Online merchandise, inventory, software, hardware, retails, wholesales etc. |
| | | Infrastructure | Roads, electricity plants, bridges, rail tracks, warehouse, residences etc. |

From equations 1, 2 and 3 –previously discussed in the introductory section– we state that the dynamic system complexity is directly proportional to the dynamic complexity of the city. If we assume a dynamic system state, $S_T$, defined as the mode or condition of a system; and which is the reciprocal of the dynamic system complexity as given by:

$$S_T = \frac{1}{\Delta S_C} \tag{4}$$

Then, it can be deduced that as

$$\lim_{k_o \to \infty} \frac{1}{\Delta S_C} = \lim_{k_o \to \infty} S_T = 0 \tag{5}$$

From (5), it can be stated that a collaborative system, consisting of different subsystems such as B2B, B2G, C2B, G2B etc., cease to exist if (or whenever) $k_o$ tends to infinity regardless of the agglomerative state of the individual complexities of the city.

## 3.2. Conceptualizing city logistics systems

The strategies for collaboration rest on the need to create and expand the semi-intangible attributes and to optimize the use of intangible attributes of a city. This can be achieved by fostering collaboration of stakeholders with efforts toward understanding the dynamic nature of individual city complexities such as the socio-cultural values of residents to demands, administrative and government policies, and activities of shippers and freight carriers as well as the environment, infrastructures and information technologies. Also, market competitions between organizations cannot always be a viable option for achieving optimum performance within the city and as such have to be minimized to a reasonable level. In addition, stakeholders through collaboration can mitigate the effect of uncertainty by eliminating all forms of pollutions, as much as possible, and making continuous improvement of their environments; for instance, by beautification and afforestation.

Evaluating these strategies might be based on how well the collaborative communities utilize the key knowledge of the axioms of uncertainty effect and the spider networks of CL. One approach for such evaluation is the use of system dynamics model (SDM) or causal loop diagram for visualizing the relationship between the elements of a CL system. Figure 6 illustrates the SDM for understanding the CL system. The link positive polarity (+) links or points from one variable parameter to another and implied that an increase (or decrease) in input variable parameter would lead to an increase (or decrease) in the output variable parameter. For instance, increasing (or decreasing) changes in administrative policies as determined by tax rates, regulations and structuring directly lead to increasing ( or decreasing) changes in shippers' activities and information technology. Likewise, link positive polarities exist between shippers and freight carriers on the basis of supply rate of goods, and also between residents and information technologies. It can be observed that an increase in the activities of freight carrier and infrastructures contributes to increasing congestions within the city system. The term "congestion" in the context of CL can be defined as space minus freights and infrastructures. An important observation is that collaboration and pollutions have link positive polarities from nearly all the variables that extend from the administrators, freight carriers, shippers and residents.

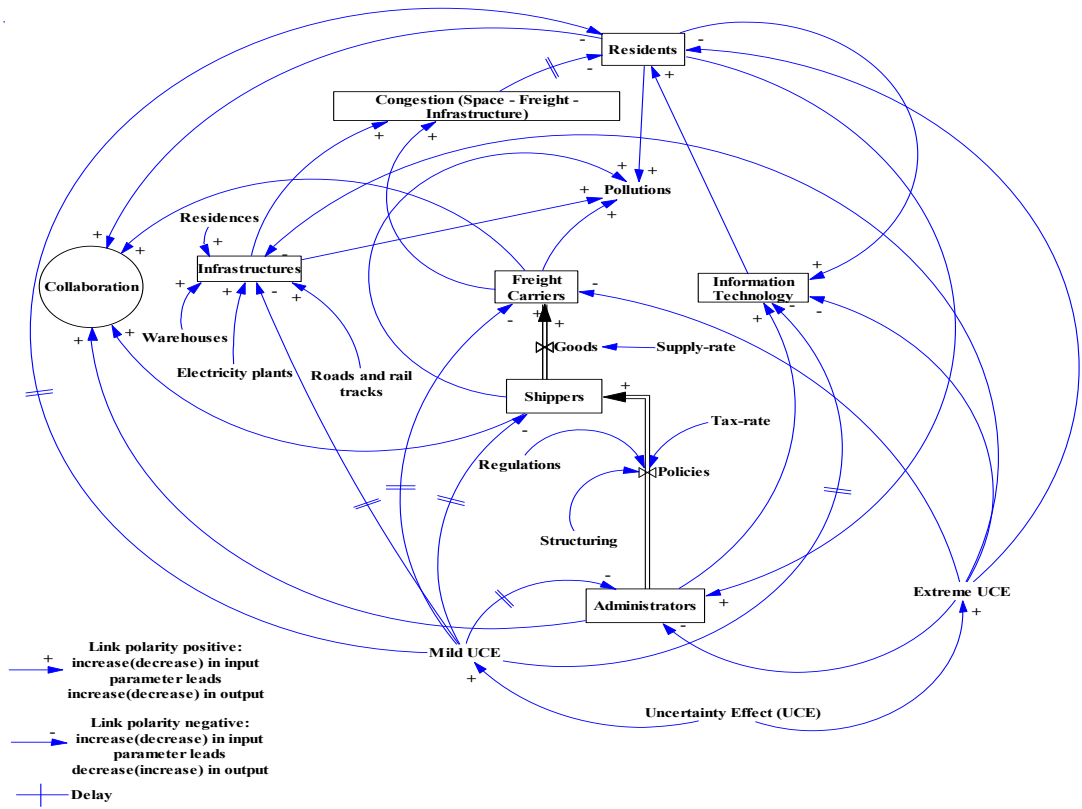

**Fig. 6: SDM for visualizing CL operations (designed with Vensim[3])**

Similarly, the link negative polarity (-) implied that an increase (or decrease) in input variable parameter would lead to a decrease (increase) in the output variable parameter. The uncertainty effect projects a link negative polarity to almost all the variables identified in the SDM. The uncertainty effect is classified as mild and extreme; and the difference between these two uncertainties has to do with the fact that there is always a delay associated with mild uncertainty effect while instantaneity is associated with extreme uncertainty effect. This means that increasing (or decreasing) mild and extreme uncertainty effect causes a decrease (or increase) in the dynamic states of the city complexities, which undergoes delay if the uncertainty effect is mild and instantaneous if the uncertainty effect is extreme. Simulating the realities of an uncertainty effect can be a daunting if not an impossible task.

---

[3] Vensim is a registered trademark of the Ventana Systems, Inc. http://www.vensim.com/

Therefore, these explanations meant that the interests of stakeholders are best assured by enhancing collaboration, as their numbers and activities increase in order for them to combat the growing challenges of environmental pollutions and the uncertainty effect.

## 3.2. Collaboration as enabler for sustainable city logistics

Collaboration strategies for CL planning can be defined as the mechanisms for achieving successful collaboration based on subsystems, such as B2B, B2C, C2C, C2B, G2B, and E2B and so on, by the stakeholders living in a city within an environment. The emergence of internet as a major medium of exchange of information, in an era of information technology, has enveloped the various subsystems to function within a larger system framework called e-commerce or e-business or –as some prefer to call it– internet collaboration or e-collaboration. Garner (1999) prefers to use the term "c-commerce" meaning collaborative commerce in describing the emergence of a new model for business applications that unfolds into subsystems.

In this research, our discussion on collaboration strategies will be limited to four key subsystems namely: business to business (B2B), business to customer (B2C), customer to customer –also known as client to client– (C2C), and customer to business (C2B). These four subsystems are often the basic and major means of doing businesses that provide mutual benefits to its stakeholders comprising of shippers, freight carriers and residents as well as administrators involved in CL operations. The administrators play the central role of an umpire that makes policies, enforces regulations and provides infrastructures and information technologies, such as internet and intranet facilities, as well as generates revenues through taxations. Table 5 illustrates these subsystems, plausible instances within the subsystem and common examples of firms and online business.

**Table 5: B2B, B2C, C2C and C2B subsystems**

| Subsystems | Plausible instance | Example |
|------------|--------------------|---------|
| **B2B** | Retailer to wholesaler, wholesaler to retailer, shippers to freight carriers, retailer to retailer, wholesaler to wholesaler, shippers to shippers, freight carriers to freight carriers. | Big marketing firms like General Motors, eBay, etc. |
| **B2C** | Retailer to customer, wholesaler to customer, shippers to customers, freight carriers to customer. | A day care business, online shopping website like Amazon, online advertising businesses like Google Ad, Yahoo Ad, etc. |
| **C2B** | Customer to retailer, customer to wholesaler, customer to shippers, customer to freight carriers. | An inventor that puts his/her invention for sale and auction to a firm, etc. |
| **C2C** | One-to-one C2C and one-to-many C2C. | Online auction, eBay etc. |

We propose two CL collaboration square (matrix) models for explaining the strategies for collaboration based on these key subsystems, namely:

1. Simplified collaboration square (SCS) model

2. Advanced collaboration square (ACS) model

### 3.3.1. Simplified collaboration square (SCS) model

The SCS model specifically represents the basic subsystems as four squares enclosed in a larger square with two common diagonal arrows pointing outward and overlapping at the center. Figure 7, illustrates the simplified collaboration square (SCS) model of B2B, B2C, C2C, and C2B. The squares in the upper horizontal row – that is, B2B and B2C– enjoy greater financial popularity when compared to the squares in the lower horizontal row – that is, C2B and C2C. Also, the squares on the left vertical column – that is, B2B and C2B– generate greater awareness and size when compared to the squares on the right vertical column – B2C and C2C.

Furthermore, the two diagonal lines with outward pointing arrows, which are Consolidation center (CC) and Social Network (SN), indicate that while B2B and C2C subsystems are adopting CC; on the other hand, C2B and B2C are adopting SN as their respective collaboration strategies. The CC and SN are fast becoming a formidable collaboration strategies being used by collaborative communities towards achieving sustainable CL operations. Typical example of CC is Wal-Mart and SN is Facebook.

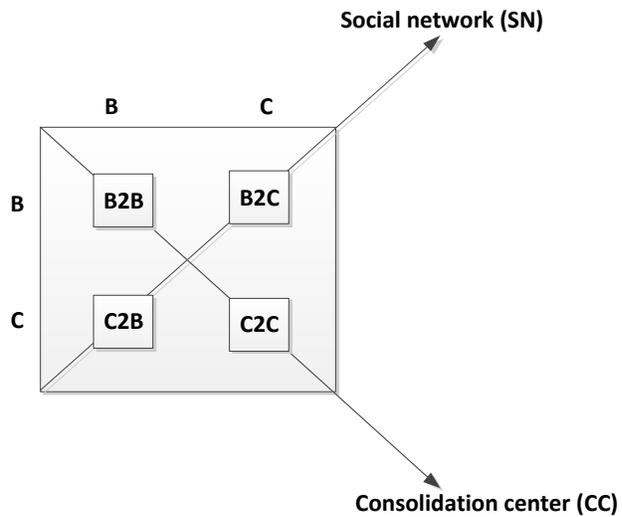

**Fig. 7: Simplified collaboration square (SCS) model of B2B, B2C, C2B, and C2C**

The highlights of Wal-Mart's 2010 financial annual report[a] estimated its net sales at $405 Billion USD and operating income of $24 Billion USD with unit counts of 8,146 worldwide. A 2010 press release from the online news website, Reuters[b] on the social networking site, Facebook, has it that "Facebook's financial performance is stronger than previously believed, as the Internet social network's explosive growth in users and advertisers boosted 2009 revenue to as much as $800 million…" One major difference between CC and SN is that while CC is visible, SS are invisible.

The advantages of collaboration in CC and SN are listed as follow:

1. They are able to achieve enormous business profits.
2. They have large business markets.
3. They help to reduce congestions in city.
4. They make commerce easy.

5. They often provide affordable services to multiple customer and businesses, simultaneously.

6. While CC can be environmental friendly, SN can be user friendly.

7. They both minimize environment imparts of stakeholders' activities that causes pollution in the city.

8. Firms involved in CC and SN often gets higher incentives from government such as higher tax returns, than their counterparts that are not into CC and SN collaborative strategies.

Second, if the dynamic environment, $\Delta E$, and uncertain effect (UCE) are to be considered as important factors that have significant influence on subsystems, as we have explained with the mathematical models of uncertainty and spider networks of CL, then the SCS model transforms to the advanced collaboration square (ACS) model of B2B, B2C, C2C, and C2B.

### 3.3.2. Advanced collaboration square (ACS) model

The ACS model unlike the SCS model has the depiction of the UCE (i.e. uncertainty effect) at the center of the two diagonal arrows, and $\Delta E$ exists within and around the dynamic system, $\Delta S_C$, as well as $\Delta A$ which is the dynamic agglomerative state of the city with exclusion of the $\Delta E$. The ACS model is illustrated in the following figure 8.

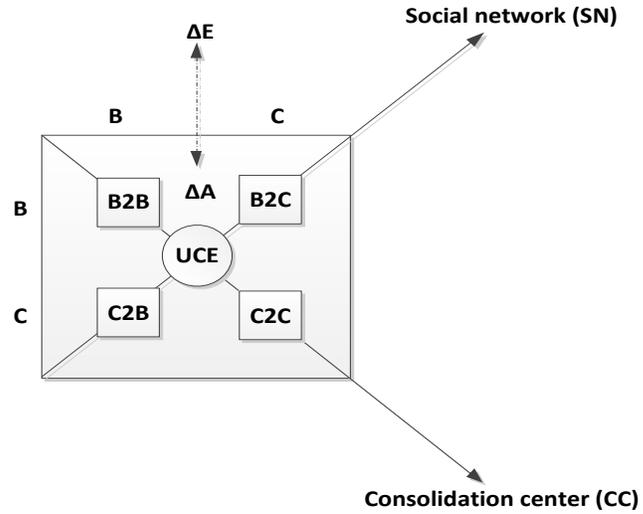

**Fig. 8: Advanced collaboration square (ACS) model for B2B, B2C, C2B, and C2C**

In addition to our earlier mathematical models, the following were further established based on the SCS and ACS CL models:

For the SCS model, it can be inferred that

$$\sum_{i=x}^{n} B_i 2\, B_i + \sum_{j=y}^{n} C_j 2\, C_j \;=\; CC \tag{6.1}$$

$$\sum_{i=x}^{n} C_i 2\, B_i + \sum_{j=y}^{n} B_j 2\, C_j \;=\; SN \tag{6.2}$$

Where *n* can vary from one to infinity, $x, y \; \epsilon \; N$ and *CC* and *SN* are assumed to be variables. Different forms could emerge from equations (6.1) and (6.2) depending on the minimum number of stakeholders involved in each of the collaborative strategies as determined by the value of $x$ and $y$.

For the ACS model, it can be inferred that

$$\Delta S_{cc} = \frac{cc}{k_o(\Delta A \cup \Delta E)} \quad\quad\quad (7.1)$$

$$\Delta S_{SN} = \frac{SN}{k_o(\Delta A \cup \Delta E)} \quad\quad\quad (7.2)$$

Where $\Delta S_{cc}$ equals the dynamic complexity of the B2B and C2C subsystems divided by the dynamic complexity of the city, $k_o(\Delta A \cup \Delta E)$. Similarly, $\Delta S_{SN}$ equals the dynamic complexity of the C2B and B2C subsystems divided by the dynamic complexity of the city, $k_o(\Delta A \cup \Delta E)$.

$$\Delta A = \Delta P_o \cup \Delta S \cup \Delta I_t \cup \Delta I \cup \Delta R \cup \Delta F_e \cup \Delta G \cup ... \quad\quad \text{(with exclusion of } \Delta E)$$

By default, and based on human levity tendency[4], some presumptions arose that:

$$k_o \cong 1, \quad\quad \Delta A \cong 0 \quad and \quad \Delta E \cong 1$$

The simplification of the mathematical equations of the ACS model satisfies (6.1) and (6.2), respectively, resulting in the SCS model. Also, it suffices to state that (7.1) and (7.2) satisfies (1) as can be seen from:

$$\Delta S_C = k_o(\Delta A \cup \Delta E)(\Delta S_{cc} + \Delta S_{SN} + \cdots) \quad\quad\quad (8)$$

---

[4] The human levity tendency (HLT) is defined as the behavioral way of assuming all is normal within a human set boundary despite contrary evidence pointing to events, elsewhere, outside the boundary.

Hence, our hypothesis that the CL system complexity will either increase or decrease due to redundancy or emergence of collaborative subsystem(s) with respect to the change in the complexity of a city.

Finally, from (8), three conditions (or trio conditionality) emerge:

1. For a city in non-chaotic situation , that is in a normal state

$$\frac{\Delta S_C}{k_o(\Delta A \cup \Delta E)} = (\Delta S_{cc} + \Delta S_{SN} + \cdots) \cong 1 \tag{9}$$

2. For a city in a near chaotic situation

$$0 < \frac{\Delta S_C}{k_o(\Delta A \cup \Delta E)} < 1 \qquad \text{- Non-approximate} \tag{10}$$

3. For a city in cataclysmic chaotic state

$$\frac{\Delta S_C}{k_o(\Delta A \cup \Delta E)} = (\Delta S_{cc} + \Delta S_{SN} + \cdots) \cong 0 \tag{11}$$

A non-chaotic situation is a normal state that city dwellers expect to have wherein there are minor accidents to no natural and man-made disasters. It is what stakeholders through collaborative planning can achieve sustainable CL operations.

A near chaotic situation would arise mainly from natural disasters of huge proportion such as a hurricane, a tsunami, a high magnitude earthquake and so on, which could engulf a city for a short time period lasting from few hours to few days.

One which should never be hoped for is that of a cataclysmic chaotic state that might result from an extremely high negative effector such as an atomic explosion, a planetoid, a black hole and so on, which would unleash uncontrollable chain reactions that could engulf a city and other cities far and near.

The approximate equalities of one and zero for the equations contained in (9) and (11) give credence to the law of conservation of mass which states that the mass of an isolated system will remain constant over time. Otherwise, if we are to assume exact and greater than equalities of one and exact equality of zero for these equations then the law of conservation of mass would be nullified with the implication that a city can be totally sustained, that is created, or destroyed.

### 3.4. Modeling and evaluating collaboration strategies

In the section, we will present solution approach for investigating collaboration strategies at two levels:

1. Macro-level

2. Micro-level

### 3.4.1. Macro-level

At the macro level, we will use the SCS model (4.3.1) to evaluate collaboration subsystems on socio-cultural characteristics($S$), economy ($E$) and environmental ($En$) impacts on the movement of goods inside the city centers under the condition of non-chaotic situation with the presumptions of human levity tendency (HLT). The SCS model avoids the mathematical complications of uncertainty effect associated with the ACS model.

Recall that for the SCS Model

$$\sum_{i=x}^{n} B_i \, 2B_i \ + \ \sum_{j=y}^{n} C_j \, 2C_j \ = CC$$

$$\sum_{i=x}^{n} C_i \, 2B_i \ + \ \sum_{j=y}^{n} B_j \, 2C_j \ = SN$$

In this case, $n = 3$ for the three instances ($S$, $E$ and $En$), such that

$$\sum_{i=1}^{3} B_i \, 2B_i = B_1 2B_1 \ + \ B_2 2B_2 + \ B_3 2B_3 = 100 \ \% \tag{12.1}$$

$$\sum_{j=1}^{3} C_j \, 2C_j \ = C_1 2C_1 \ + \ C_2 2C_2 \ + \ C_3 2C_3 = 100 \ \% \tag{12.2}$$

$$\sum_{i=1}^{3} C_i \, 2B_i \ = C_1 2B_1 \ + \ C_2 2B_2 \ + \ C_3 2B_3 = 100 \ \% \tag{12.3}$$

$$\sum_{j=1}^{3} B_j \, 2C_j \ = B_1 2C_1 \ + \ B_2 2C_2 \ + \ B_3 2C_3 = 100 \ \% \tag{12.4}$$

Figure 9 provides a diagrammatic representation of (12.1).

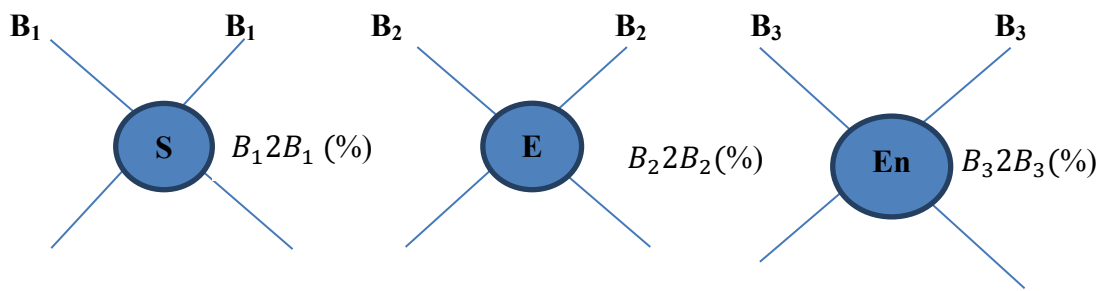

**Fig. 9: Diagrammatic representation of equation (12.1)**

Where $B_1 2B_1$, $B_2 2B_2$, $B_3 2B_3$ $\epsilon + R$ are the percentage weight randomly assigned to each of the collaboration intent on socio-cultural characteristics *(S)*, economy (*E*) and environment (*En*) impart.

Similar representation can be made for equations (12.2) to (12.4). And, taking

- Collaboration:   Plus (+)

- Non-collaboration or competition: Minus (-)

- Undecided: 0

(See Figure 10)

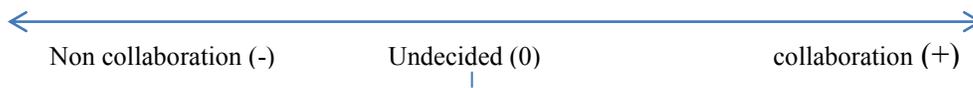

Non collaboration (-)          Undecided (0)                    collaboration (+)

**Fig. 10: Line graph for weighted collaboration intent**

The following test case illustrates the application of the SCS model. Table 6 discusses the conclusions drawn from the evaluation of this test case. Many other cases can emerge totaling $2^n \times 2^n = 2^{2n}$, where if $n = 3$ for the three instances which are the *S*, *E*, and *En* then the overall test cases should be sixty-four.

**<u>Illustration:</u>**

Let us consider the following four options are evaluated namely B2B, B2C, C2B, and C2C. The descriptions of the four options are presented in the Table 6.

| | B | | | C | | | Collaboration matrix | | | Weight (%) | |
|---|---|---|---|---|---|---|---|---|---|---|---|
| | S (+) | E (-) | En (+) | S (-) | E (-) | En (+) | | | | | |
| **B** | 10 | 40 | 50 | 20 | 10 | 70 | **SN** (0 | -50 | 110) | **SN** (60) | |
| **C** | 20 | 40 | 40 | 30 | 40 | 30 | **CC** (-20 | -80 | 80) | **CC** (-20) | |

**Table 6: Evaluation of test case**

| Option | Description of SCS Model | Conclusion on $SS$ and $CC$ |
|---|---|---|
| 1 | C2B has weighted collaboration intent of 20% for $S$, non-collaboration intent of 40% for $E$ and 40% collaboration intent for $En$, *i.e.* $(20, -40, 40)$. | Weighted collaboration intent increase to 60% for $SN$, *i.e.* $SN[\,(20, -40, 40)$ |
| | B2C has weighted non-collaboration intent of 20% for $S$, non-collaboration intent of 10% for $E$ and 70% collaboration intent for $En$, *i.e.* $(20, -10, 70)$. | $+ \ (-20, -10, 70)]$ $= SN(0, -50, 110)$ $= SN(60)$ |
| 2 | B2B has weighted collaboration intent of 10% for $S$, non-collaboration intent of 40% for $E$ and 50% collaboration intent for $En$, *,i.e.* $(10, -40, 50)$. | Weighted collaboration intent decrease to 20% for $CC$, *i.e.* $CC[\,(10, -40, 50)$ |
| | C2C has weight non-collaboration intent of 30% for $S$, non-collaboration intent of 40% for $E$ and 30% collaboration intent for $En$, *i.e.* $(-30, -40, 30)$. | $+ \ (-30, -40, 30)]$ $=CC(-20, -80, 80)$ $=CC(-20)$ |

Based on the results of Table 6, we have been able to show how to derive the weighted collaboration intent for SN and CC based on the four subsystems and the conclusions that can be made.

### 3.4.2. Micro-level

At the micro level, we will use the results obtained from the macro level (SCS model) for consolidation centers, CC, that emerge from B2B and C2C and then perform planning at the operational level. Models 2-4 are extended version of models proposed by Awasthi and Proth (2006) for investigating CL decision making. This work will be accomplished through the following models:

1. SCS (input filter) model

2. Goods to vehicle assignment model

3. Vehicle routing and scheduling

4. Environment impact assessment model

These conjoined models are called the CL operational level model. These four models are inter-related to each other via input and output variables as illustrated in Figure 11. In the following subsections, these steps are described in detail.

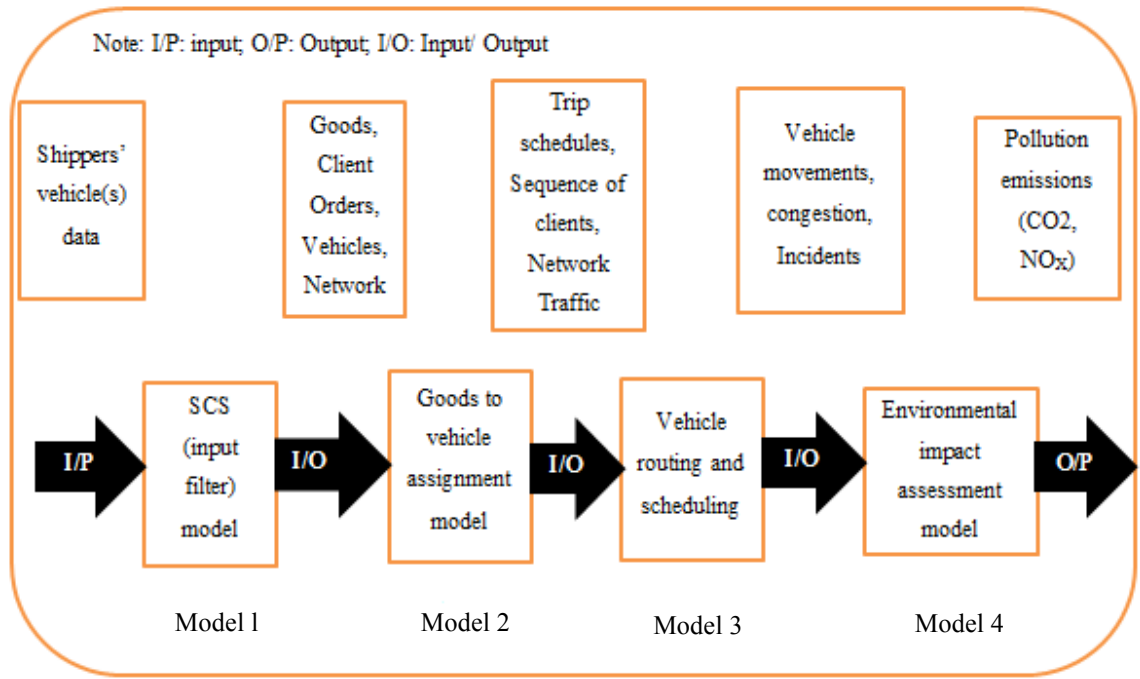

**Fig. 11: Operational level model**

### 3.4.2.1. SCS (input filtering) model

The SCS model performs filtering of goods vehicles based on the satisfaction of the constraints for compliance with usage in the city.

By this model, the following constraints can be established:

1. Constraints C1: City restriction on vehicle size $\leq$ maximum admissible vehicle size $\times$ collaborative intent on social-cultural characteristics

2. Constraint C2: Gains restriction from distribution (for plausible taxation purpose) $\geq$ maximum admissible net profit $\times$ collaborative intent on economy

3. Constraints C3: Environment restriction on pollutants $\geq$ maximum admissible pollutant level $\times$ Collaborative intent on environment

### 3.4.2.2. Goods to vehicle assignment model

The Goods to vehicle assignment model allocates goods to vehicles respecting the load capacity of the vehicles. The vehicles are assumed to be trucks of varying capacities with permissible emission factors.

### 3.4.2.3. Goods distribution model

The Goods distribution model performs the scheduling and routing of goods vehicles from the source of destination. The routing and scheduling decisions for vehicles are done respecting the delivery time window (VRPTW) and in line with the principle of TSP.

### 3.4.2.4. Environmental impact assessment model

The environmental impact assessment model evaluates the impact of goods movement on the city environment. As shown in Figure 11, plausible considerations for the environmental impact assessment model are the vehicle movements, congestion, incidents and other variables that form the bases for evaluating the pollution and emissions ($CO_2$, $NO_x$) level for goods vehicles in the city.

### 3.5. Complementarity of macro-level and micro-level models

The complementarity of the macro-level and micro-level models proposed above to address collaboration planning problems is to find a measure of the intents of decision makers to collaborate at the macro level, for example from business-to-customer and vice-versa, the outcome of which is the determination of their weighted collaborative intents that influence the activities of city operators at the micro-level of city logistics. Importantly, the SCS (macro-level) model provide a method for evaluating socio-cultural characteristics, economy and environmental impacts, and so on, deemed vital to the performance of city logistics systems within the framework of the basic and major subsystems, B2B, B2C, C2B, and C2C, that are (rapidly) unifying into either consolidation centers or social networks; while, the micro-level models finds a basis for the application of the macro level model at the level of operations of the city logistics operators.

Specifically, this research focus is on the use of the macro-level models with respect to consolidation centers at the micro-level. The visibility of consolidation centers, unlike the invisibility of social network, makes it easier to place into theoretical and practical perspectives. But, we envisage that future work could find the usability of the macro-level models with regards to social networks at the micro-level of city logistics operations.

# Chapter 4:

# Numerical analysis

## 4.1. Overview

This numerical analysis is done at the macro-level with the collaboration model and at the micro- level with the operation level model.

### 4.1.1. Collaboration square model

We shall answer the question of "Which test case is best suited for collaboration from the B2B, B2C, C2B, and C2C subsystems in the formation of *SN* and *CC* considering the collaboration matrix and weighted collaboration intent (%) of the test cases?"

To answer this, five sample test cases are presented in the Table 7. It can be seen that test case 5 is best suited for collaboration for the subsystems based on the weighted collaboration intent of *SN* and *CC* with the highest global optimum for the five test cases under consideration. Also, test cases 1 and 4 can be considered satisfactory for collaboration towards *SN* and *CC* since their respective weighted collaborative intent are above zero percent. Also, it can be concluded that test case 2 may be unsuitable for *SN* with C2B and B2C subsystems due to a decline towards non-collaboration as can be seen by the negative weighted collaborative intent; while, the result showed indecision with B2B and C2C for collaboration for *CC*. For test case 3, a conclusion can be reached that due to the negative weighted collaborative intents for the *SN* and CC that non-collaboration exists for the subsystems.

## Test sample 1

*Case 1:*

|  | B |  |  | C |  |  | Collaboration matrix | | | Weight (%) | |
|---|---|---|---|---|---|---|---|---|---|---|---|
|  | S (+) | E (-) | En (+) | S (+) | E (-) | En (+) |  |  |  |  |  |
| B | 10 | 40 | 50 | 20 | 10 | 70 | SN (40 | -50 | 110) | SN (100) | |
| C | 20 | 40 | 40 | 30 | 40 | 30 | CC (40 | - 80 | 80) | CC (40) | |

Case 2:

|  | B |  |  | C |  |  | Collaboration matrix | | | Weight (%) | |
|---|---|---|---|---|---|---|---|---|---|---|---|
|  | S (+) | E (-) | En (+) | S (-) | E (+) | En (-) |  |  |  |  |  |
| B | 10 | 40 | 50 | 20 | 10 | 70 | SN(0 | -30 | -30) | SN(-60) | |
| C | 20 | 40 | 40 | 30 | 40 | 30 | CC (-20 | 0 | 20) | CC (0) | |

Case 3:

|  | B |  |  | C |  |  | Collaboration matrix | | | Weight (%) | |
|---|---|---|---|---|---|---|---|---|---|---|---|
|  | S (+) | E (+) | En (-) | S (-) | E (+) | En (-) |  |  |  |  |  |
| B | 10 | 40 | 50 | 20 | 10 | 70 | SN (0 | 50 | -110) | SN (-60) | |
| C | 20 | 40 | 40 | 30 | 40 | 30 | CC (-20 | 80 | -80) | CC (-20) | |

Case 4:

|  | B |  |  | C |  |  | Collaboration matrix | | | Weight (%) | |
|---|---|---|---|---|---|---|---|---|---|---|---|
|  | S (+) | E (+) | En (-) | S (-) | E (+) | En (+) |  |  |  |  |  |
| B | 10 | 40 | 50 | 20 | 10 | 70 | SN (0 | 50 | -30) | SN (20) | |
| C | 20 | 40 | 40 | 30 | 40 | 30 | CC (-20 | 80 | -20) | CC (-40) | |

Case 5:

|  | B |  |  | C |  |  | Collaboration matrix | | | Weight (%) | |
|---|---|---|---|---|---|---|---|---|---|---|---|
|  | S (+) | E (-) | En (+) | S (+) | E (+) | En (+) |  |  |  |  |  |
| B | 10 | 40 | 50 | 20 | 10 | 70 | SN (40 | -30 | 110) | SN (120) | |
| C | 20 | 40 | 40 | 30 | 40 | 30 | CC (40 | 0 | 80) | CC (120) | |

**Table 7: Analyses of the five test cases –ref. Test sample 1**

| Test Cases | Description of SCS Model | Inference on *SS* and *CC* |
|---|---|---|
| 1 | C2B has weighted collaboration intent of 20% for *S*, non-collaboration intent of 40% for *E* and 40% collaboration intent for *En, i.e.* $(20, -40, 40)$. | Weighted collaboration intent increase to100% for *SN, i.e.* $SN[\,(20, -40, 40)$ $+ (20, -10, 70)]$ |
| | B2C has weighted collaboration intent of 20% for *S*, non-collaboration intent of 10% for *E* and 70% collaboration intent for *En, i.e.* $(20, -10, 70)$. | $= SN(40, -50, 110)$ $= SN(100)$ |
| | B2B has weighted collaboration intent of 10% for *S*, non-collaboration intent of 40% for *E* and 50% collaboration intent for *En, ,i.e.* $(10, -40, 50)$. | Weighted collaboration intent increase to 40% for *CC, i.e.* $CC[\,(10, -40, 50)$ $+ (30, -40, 30)]$ |
| | C2C has weight collaboration intent of 30% for *S*, non-collaboration intent of 40% for *E* and 30% collaboration intent for *En, i.e.* $(30, -40, 30)$. | $= CC(40, -80, 80)$ $= CC(40)$ |
| 2 | C2B has weighted collaboration intent of 20% for *S*, non-collaboration intent of 40% for *E* and 40% collaboration intent for *En, i.e.* $(20, -40, 40)$. | Weighted non-collaboration intent decrease to 60% for *SN, i.e.* $SN[\,(20, -40, 40)$ |
| | B2C has weighted non-collaboration intent of 20% for *S*, collaboration intent of 10% for *E* and 70% non-collaboration intent for *En, i.e.* $(20, -10, 70)$. | $+ (-20, 10, -70)]$ $= SN(0, -30, -30)$ $= SN(-60)$ |
| | B2B has weighted collaboration intent of 10% for *S*, non-collaboration intent of 40% for *E* and 50% collaboration intent for *En, i.e.* $(10, -40, 50)$. | Undecided (0%) weighted collaboration intent for CC, *i.e.* $CC[\,(10, -40, 50)$ |
| | C2C has weight non-collaboration intent of 30% for *S*, collaboration intent of 40% for *E* and 30% non-collaboration intent for *En, i.e.* $(-30, 40, -30)$. | $+ (-30, 40, -30)]$ $= CC(-20, 0, 20)$ $= CC(0)$ |
| 3 | C2B has weighted collaboration intent of 20% for *S*, collaboration intent of 40% for *E* and 40% non-collaboration intent for *En, i.e.* $(20, 40, -40)$. | Weighted non-collaboration intent decrease to 60% for *SN, i.e.* $SN[\,(20, 40, -40)$ |
| | B2C has weighted non-collaboration intent of 20% for *S*, collaboration intent of 10% for *E* and 70% non-collaboration intent for *En, i.e.* $(-20, 10, -70)$. | $+ (-20, 10, -70)]$ $= SN(0, 50, -110)$ $= SN(-60)$ |
|  | B2B has weighted collaboration intent of 10% for $S$, collaboration intent of 40% for $E$ and 50% non-collaboration intent for $En$, *i.e.* $(10,40,-50)$. | Weighted non-collaboration intent decrease to 20% for $CC$, *i.e.* $CC[\,(10,40,-50)$ $+(-30,40,-30)]$ $=CC(-20,80,-80)$ $=CC(-20)$ |
|  | C2C has weight non-collaboration intent of 30% for $S$, collaboration intent of 40% for $E$ and 30% non-collaboration intent for $En$, *i.e.* $(-30,40,-30)$ | |
| 4 | C2B has weighted collaboration intent of 20% for $S$, collaboration intent of 40% for $E$ and 40% non-collaboration intent for $En$, *i.e.* $(20,40,-40)$. | Weighted collaboration intent increase to 80% for $SN$, *i.e.* $SN[\,(20,40,-40)$ $+(-20,10,70)]$ $=SN(0,50,30)$ $=SN(80)$ |
|  | B2C has weighted non-collaboration intent of 20% for $S$, collaboration intent of 10% for $E$ and 70% collaboration intent for $En$, *i.e.* $(-20,10,70)$ | |
|  | B2B has weighted collaboration intent of 10% for $S$, collaboration intent of 40% for $E$ and 50% non-collaboration intent for $En$, *i.e.* $(10,40,-50)$. | Weighted collaboration intent increase to 40% for $CC$, *i.e.* $CC[\,(10,40,-50)$ $+(-30,40,30)$ $=CC(-20,80,-20)$ $=CC(40)$ |
|  | C2C has weight non-collaboration intent of 30% for $S$, collaboration intent of 40% for $E$ and 30% collaboration intent for $En$, *i.e.* $(-30,40,30)$. | |
| 5 | C2B has weighted collaboration intent of 20% for $S$, non-collaboration intent of 40% for $E$ and 40% collaboration intent for $En$ *i.e.* $(20,-40,40)$. | Weighted collaboration intent increase to 120% for $SN$ *i.e.* $SN[\,(20,-40,40)$ $+(20,10,70)]$ $=SN(40,-30,110)$ $=SN(120)$ |
|  | B2C has weighted collaboration intent of 20% for $S$, collaboration intent of 10% for $E$ and 70% collaboration intent for $En$ *i.e.* $(20,10,70)$. | |
|  | B2B has weighted collaboration intent of 10% for $S$, collaboration intent of 40% for $E$ and 50% non-collaboration intent for $En$, *i.e.* $(10,-40,50)$. | Weighted collaboration intent increase to 120% for $CC$ *i.e.* $CC[\,(10,-40,50)+$ $(30,40,30)$ $=CC(40,0,80)$ $=CC(120)$ |
|  | C2C has weight collaboration intent of 30% for $S$, collaboration intent of 40% for $E$ and 30% collaboration intent for $En$, *i.e.* $(30,40,30)$. | |

Figure 12, with reference to Test sample 1, illustrate the line plots of *SN* and *CC* weighted collaborative intents for the five test cases with their respective data points and the maximum and minimum points.

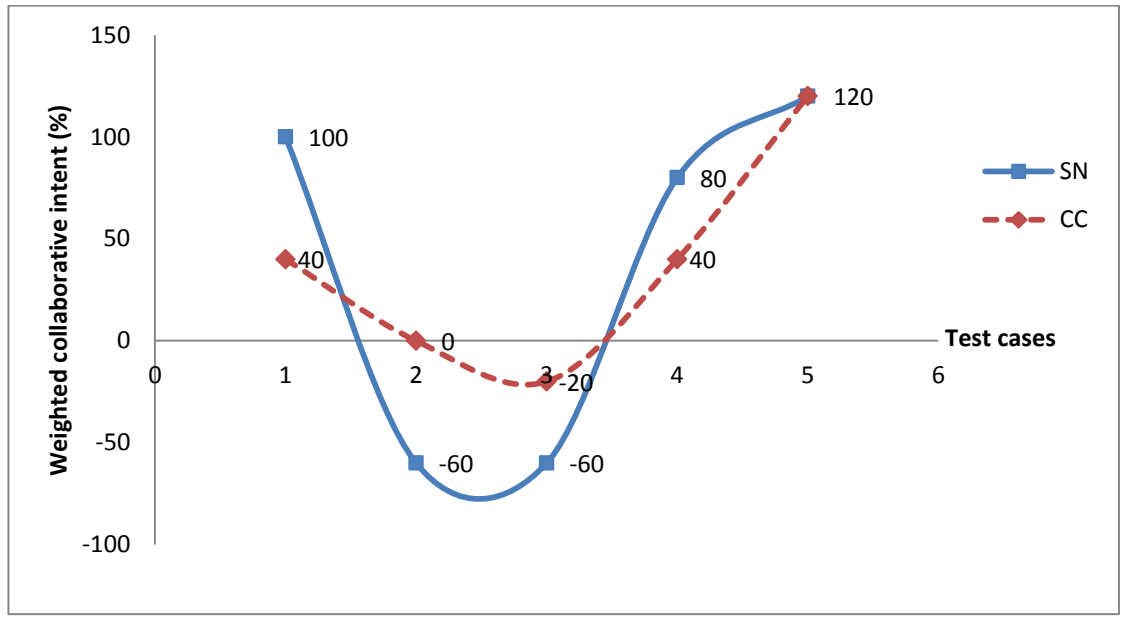

**Fig. 12: Line plots for SN and CC weighted collaborative intents versus test cases**

### 4.1.2. Operational planning model

Let us assume that there are six shippers each serving six clients in a given city area. Table 8 presents the clients demand orders for the six shippers: S1, S2, S3, S4, S4 and S6. Table 9.1 to 9.6 presents the trucks available to the shippers and Table 10 describes the emission standards for passenger cars and light-duty trucks. Our goal is to evaluate collaboration possibilities between them and decide which collaboration case assist to reduce vehicular emission respecting VRSTW in accordance to the TSP.

To perform this study, we will use the use the operational planning model (micro level) to assess collaboration possibilities between the various shippers.

**Table 8:  Clients demand orders for shippers S1, S2, S3, S4, S5 and S6**

| Shippers | Clients | No of packets to be delivered(N) | Size of the Packet(S) | Time windows | Quantity to be delivered= (N*S) |
|---|---|---|---|---|---|
| S1 | C1 | 10 | 20 | 9am-10am | 200 |
|  | C2 | 10 | 2 | 10am-11am | 20 |
|  | C3 | 10 | 10 | 11am-12pm | 100 |
|  | C4 | 10 | 6 | 12pm-1pm | 60 |
| S2 | C5 | 10 | 10 | 9am-10am | 100 |
|  | C6 | 10 | 5 | 9am-10am | 50 |
|  | C7 | 20 | 10 | 9am-10am | 200 |
|  | C8 | 5 | 10 | 9am-12pm | 50 |
| S3 | C6 | 10 | 10 | 9am-10am | 100 |
|  | C7 | 10 | 5 | 9am-10am | 50 |
|  | C8 | 25 | 2 | 9am-12pm | 50 |
|  | C10 | 10 | 10 | 9am-10am | 100 |
| S4 | C7 | 10 | 5 | 9am-10am | 50 |
|  | C8 | 10 | 10 | 9am-12pm | 100 |
|  | C9 | 1 | 30 | 9am-10am | 30 |
|  | C11 | 1 | 30 | 9am-10am | 30 |
| S5 | C12 | 10 | 5 | 9am-10am | 50 |
|  | C13 | 20 | 4 | 9am-11am | 80 |
|  | C14 | 5 | 10 | 9am-11am | 50 |
|  | C15 | 10 | 20 | 9am-11am | 200 |
| S6 | C1 | 20 | 10 | 9am-10am | 200 |
|  | C2 | 10 | 2 | 10am-11am | 20 |
|  | C3 | 10 | 10 | 11am-12pm | 100 |
|  | C4 | 10 | 6 | 12pm-1pm | 60 |

**Table 9.1: Trucks available with S1**

| Trucks | Load capacity (kg) | Vehicle size(tons) | Emission Factor |
|---|---|---|---|
| T1 | 100 | 200 | E1 |
| T2 | 200 | 240 | E2 |

**Table 9.2: Trucks available with S2**

| Trucks | Load capacity (kg) | Vehicle size(tons) | Emission Factor |
|---|---|---|---|
| T3 | 100 | 330 | E2 |
| T4 | 200 | 200 | E1 |

**Table 9.3 Trucks available with S3**

| Trucks | Load capacity (kg) | Vehicle size(tons) | Emission Factor |
|---|---|---|---|
| T5 | 100 | 200 | E1 |
| T6 | 200 | 300 | E2 |

**Table 9.4 Trucks available with S4**

| Trucks | Load capacity (kg) | Vehicle size(tons) | Emission Factor |
|---|---|---|---|
| T7 | 50 | 150 | E1 |
| T8 | 100 | 150 | E2 |

**Table 9.5: Trucks available with S5**

| Trucks | Load capacity (kg) | Vehicle size(tons) | Emission Factor |
|---|---|---|---|
| T9 | 80 | 400 | 1.1E1 |
| T10 | 200 | 300 | 2E2 |

**Table 9.6: Trucks available with S6**

| Trucks | Load capacity (kg) | Vehicle size(tons) | Emission Factor |
|---|---|---|---|
| T1 | 100 | 200 | E1 |
| T2 | 200 | 240 | E2 |

**Table 10: Emission standard (Source: National LEV program)**

EPA Tier 1 Emission Standards for Passenger Cars and Light-Duty Trucks, FTP 75, g/mi

| Category | 50,000 miles/5 years | | | | | | 100,000 miles/10 years[1] | | | | | |
|---|---|---|---|---|---|---|---|---|---|---|---|---|
| | THC | NMHC | CO | $NO_X$ Diesel | $NO_X$ Gasoline | PM | THC | NMHC | CO | $NO_X$ Diesel | $NO_X$ Gasoline | PM |
| Passengers cars | 0.41 | 0.25 | 3.4 | 1.0 | 0.4 | 0.08 | - | 0.31 | 4.2 | 1.25 | 0.6 | 1.0 |
| LLDT, LVW <3, 750 lbs | - | 0.25 | 3.4 | 1.0 | 0.4 | 0.08 ⁻ | 0.80 | 0.31 | 4.2 | 1.25 | 0.6 | 0.10 |
| LLDT, LVW >3, 750 lbs | - | 0.32 | 4.4 | - | 0.7 | 0.08 ⁻ | 0.80 | 0.40 | 5.5 | 0.97 | 0.97 | 0.10 |
| HLDT, ALVW <5,750 lbs | 0.32 | - | 4.4 | - | 0.7 | - ⁻ | 0.80 | 0.46 | 6.4 | 0.98 | 0.98 | 0.10 |
| HLDT, ALVW >5,750 lbs | 0.9 | - | 5.0 | - | 1.1 | - ⁻ | 0.80 | 0.56 | 7.3 | 1.53 | 1.53 | 0.12 |

1–Useful life 120,000 miles/11 years for all HLDT standards and for all THC standards for LDT

Abbreviations:

LVW –Loaded vehicle weight (curb weight + 300lbs)

ALVW – Adjusted LVW ( the numerical average of the curb weight and the GVWR)

LLDT – Light light-duty truck ( below 6,000 lbs GVWR)

HDLT – Heavy light-duty truck( above 6,000lbs GVWR)

***Step 1: SCS Initial input filtering (Vehicle usage compliance model)***

In this step, we carried out the filtering of the shippers' vehicles for compliance with restrictions on movement of shippers' vehicles in the city centers. To achieve this, from the Test sample 2 we selected Case 5 having collaborative matrix (40, 0, 80) due to its high weighted collaborative intent of 120% for CC; then, for administrative purpose, the likely constraints that can be set are:

C1: City restriction on vehicle size $\leq$ 600tons $\times$ 40%

$$\leq 240\text{tons}$$

C2: Gains on distribution (for plausible taxation purpose) $\geq$ \$5000 $\times$ 0%

$$\geq \$0$$

C3: Environment restriction on emission factor $\geq$ E1 $\times$ 80% *OR* E2 $\times$ 80%

The constraints can represent the admissible optimum values taking into consideration the weighted collaboration intents of data gotten from city logistics operators. Based on the constraints, we can say that constraint C1 is fully satisfied by shippers S1,S6 and S4 since the vehicles size for S1,S6 with trucks T1=200 (<240tons), and T2 =240tons; and for S4, T7=150tons and T8=150tons (both less than 240tons); the constraint is partially satisfied by S2 and S3 since the vehicle size for S2 trucks T3=330 (> 240tons) and T4=200 (<240tons) while, for S3, vehicle size for it trucks T5=200 (<240tons) and T6=300 (>240tons); but, S5 failed to satisfy the constraints on vehicle size since T9=400 (>240tons) and T10=300 ( > 240tons). The constraints C2 and C3 are met by all the shippers S1, S2, S3, S4 and S5 since all of them satisfied the restrictions set by these constraints. A necessary condition for the Goods-to-vehicle assignments model is that the trucks must satisfy all the three constraints C1, C2 and C3.  Table 11 summarizes the outcome of this filtering. It can be seen that the shippers S1, S2, S3, S4 and S6 fully satisfied the requirements for usage of their vehicles inside the city centers; with an assumption that they all have profiting businesses.

**Table 11: Input filtering for Shippers' vehicles**

| Shippers | Vehicles/Trucks | Constraints | | | Accept/Reject trucks | Constraints inference |
|---|---|---|---|---|---|---|
| | | C1 | C2 | C3 | | |
| S1 | T1 | √ | √ | √ | Accept | Fully satisfied |
| | T2 | √ | √ | √ | Accept | Fully satisfied |
| S2 | T3 | × | √ | √ | Reject | Partially satisfied |
| | T4 | √ | √ | √ | Accept | Fully satisfied |
| S3 | T5 | √ | √ | √ | Accept | Fully satisfied |
| | T6 | × | √ | √ | Reject | Partially satisfied |
| S4 | T7 | √ | √ | √ | Accept | Fully satisfied |
| | T8 | √ | √ | √ | Accept | Fully satisfied |
| S5 | T9 | × | √ | √ | Reject | Partially satisfied |
| | T10 | × | √ | √ | Reject | Partially satisfied |
| S6 | T1 | √ | √ | √ | Accept | Fully satisfied |
| | T2 | √ | √ | √ | Accept | Fully satisfied |

√ : Satisfied

× : Not satisfied

***Step 2: Assignment of goods to vehicle (Goods to vehicle assignment model)***

In this step, we assign the goods to the vehicles of shippers that had fully satisfied the rules set by the constraints on use of goods vehicles in the city. In order to fulfill the demands of clients itemized in Table 8, the possible assignments of goods to vehicles for the shippers S1, S2, S3 and S4 are provided in Table 12.1 to 12.4. For example, for S1 (same as with S6) with two vehicles: truck T1 and truck T2, as can be seen in Table 12.1, in Case 1.1 we assign 200kg (=10 × 20kg) to T2 for client C1, 20kg (10 × 2kg) to T1 for client C2, 100kg (=10 ×10kg) to T1 for client C3, and then 80kg (=20kg +60kg) to T1 for client C4. We can see that for this Case 1.1, that with this combination, three trips are made, two by T1 and one by T2. Similar explanations can be made for the loading of trucks for the other test cases by the shippers in which different possible combinations for loading of trucks can be derived. However, since the goal of this step is to ensure that maximum capacity utilization of vehicles takes place, we can say that for shippers S1 and S6, Case 1.7 and Case 1.8 have the minimum number of vehicle trips taking into consideration the order upon which the goods have been assigned to trucks based on their load capacity:

(Case 1.7, Case 1.8) > (Case 1.1, Case 1.2, Case 1.3, Case 1.4, Case 1.5, and Case 1.6) > (Case 1.9, Case 1.10, Case 1.11)

Also, for shippers S2 with two trucks T3 and T4, we can recall from the filtering carried out in the previous step, that truck T3 was rejected due to failure to meet the constraint on vehicle size restriction. Hence S2 can only use truck T4 (maximum capacity = 200kg) to supply its clients. As can be seen in Table 12.2 the maximum capacity for vehicle utilization can best be achieved with two trips by allotting 200kg (20 × 10) to T4 for first trip and 200kg (100 + 50 +500) for the second trip or vise-versa, that is 200kg (100 + 50 +500) for first trip and 200kg (20 × 10) for second trip by T4.

(Case 2.1 = Case 2.2)

Likewise, shipper S3 has a single truck T5 as we can observe in Table 12.3 and the possible allocations of packets to truck T5.

(Case 3.1, Case 3.2, Case 3.3) > (Case 3.4, Case 3.5)

Therefore, for maximum capacity utilization of the truck, we can say that three trips can be made by T5. The disadvantage of a shipper using a single truck to meet the demands of its clients would be inefficiency due to time delay should the single truck develop a major fault that results in breakdown.

Finally, for shippers S4 with two trucks T7 (maximum capacity= 50) and T8 (maximum capacity =100) , we observe from Table 12.3 that with Case 4.5 and Case 4.6 two trips can be made taking into account the maximum capacity utilization of the vehicle within the time window as it can be observe that:

(Case 4.5, Case 4.6) > (Case 4.1, Case 4.2, Case 4.3, Case 4.4)

The data of the load case numbers having minimum trip lengths respecting minimum capacity utilization of vehicles as determined by the possible allocation of packets to trucks served are used in the second step of routing and scheduling of vehicles.

**Table 12.1: Possible allocations of packets to trucks by S1 and S6**

| Load case number | Trucks with their loads | | | | Total number of vehicle trips |
|---|---|---|---|---|---|
| Case 1.1 | T1(100) | T2(200) | T1(20 + 60) | | 3 |
| Case 1.2 | T1(20+60) | T2(200) | T1(100) | | 3 |
| Case 1.3 | T1(100) | T1(100+ 60) | T2(200) | | 3 |
| Case 1.4 | T2(100 + 20 +60) | T1(100) | T1(100) | | 3 |
| Case 1.5 | T1(100) | T2(100 + 20 +60) | T1(100) | | 3 |
| Case 1.6 | T1(100) | T1(100) | T2(100 + 20 +60) | | 3 |
| Case 1.7 | T2(200) | T2(100+60+ 20) | | | 2 |
| Case 1.8 | T2(100+60+ 20) | T2(200) | | | 2 |
| Case 1.9 | T1(20 + 60) | T1(100) | T1(100) | T1(100) | 4 |
| Case 1.10 | T1(100) | T1(20 + 60) | T1(100) | T1(100) | 4 |
| Case 1.11 | T1(100) | T1(100) | T1(20 + 60) | T1(100) | 4 |
| Case 1.12 | T1(100) | T1(100) | T1(100) | T1(20 + 60) | 4 |

**Table 12.2: Possible allocations of packets to truck T4 by S2**

| Load case number | Trucks with their loads | | Total number of vehicle trips |
|---|---|---|---|
| Case 2.1 | T4(200) | T4(100 + 50 +50) | 2 |
| Case 2.2 | T4(100 + 50 +50) | T4(200) | 2 |

**Table 12.3: Possible allocations of packets to truck T5 by S3**

| Load case number | Trucks with their loads | | | | Total number of vehicle trips |
|---|---|---|---|---|---|
| Case 3.1 | T5(50 + 50) | T5(100) | T5(100) | | 3 |
| Case 3.2 | T5(100) | T5(50 + 50) | T5(100) | | 3 |
| Case 3.3 | T5(100) | T5(100) | T5(50 + 50) | | 3 |
| Case 3.4 | T5(100) | T5(100) | T5(50) | T5(50) | 4 |
| Case 3.5 | T5(50) | T5(50) | T5(100) | T5(100) | 4 |

**Table 12.4: Possible allocations of packets to truck T7 and T8 by S4**

| Load case number | Trucks with their loads | | | | Total number of vehicle trips |
|---|---|---|---|---|---|
| Case 4.1 | T7(50) | T8(100) | T7(30) | T7(30) | 4 |
| Case 4.2 | T8(100) | T7(30) | T7(50) | T7(30) | 4 |
| Case 4.3 | T7(30) | T7(30) | T8(100) | T7(50) | 4 |
| Case 4.4 | T7(30) | T8(100) | T7(30) | T7(50) | 4 |
| Case 4.5 | T8(100) | T8(50 + 30 +30) | | | 2 |
| Case 4.6 | T8(50 + 30 +30) | T8(100) | | | 2 |

***Step 3: Vehicle routing and scheduling (Goods distribution model)***

In this step, we perform the routing and scheduling of goods vehicles (or loaded trucks) with the minimum number of vehicle trips respecting delivery time windows in the principle of TSP. For instance, for shippers S1 and S2(see Table 12.1), each with two similar trucks (T1 and T4) we select Case 1.7 and Case 1.8 with minimum number of vehicle trips of two; for shipper S2 with truck T4 (see Table 12.2), we select Case 2.1 and Case 2.2 with minimum number of vehicle trips of two; and, for shipper S3 with truck T5(see Table 12.3), we select Case 3.1 , Case 3.2 and Case 3.3 with minimum number of vehicle trips of three; while, for shippers S4 with trucks T1 and T4(see Table 12.4), we select Case 4.5 and 4.6 with minimum number of vehicle trips of two.

Furthermore, we develop a network diagram as shown in Figure 13 to show the locations of shippers and clients and the possible routes that can be used by the shippers in delivery of goods to clients. In the figure, the link between shippers are depicted with dashed curvy lines to indicate the likelihood of collaboration between them; while, the link between shippers and clients are depicted with thin curvy lines to represent the route connecting the locations of shippers to the clients; and the links between clients are depicted with straight thin lines to represent the route connecting the locations of clients. Also, assigned to the routes are numbers that represent the triplength. The vertical "divisor" rule indicates that no connections exist between the city logistic operators located on the left side of the divisor and those located on the right side of the divisor though we assume that their activities occurred within the same city.

Table 13 presents the different schedules and possible routes for the goods vehicles for the cases. It can observe from the table that three collaboration types, namely: no collaboration (NC), partial collaboration (PC) and full collaboration (FC) exist between shippers. Given the assumption that the routes followed by the vehicles conform to the principle of TSP with goods delivered to clients in minimum time respecting delivery time windows; we can say that PC and FC of shippers satisfy these conditions, since the total triplength for PC and FC are lesser than that of NC between shippers.

In numerical terms, it can be stated that:

Total trip lengths of FC + Total trip lengths of PC < Total trip lengths of NC

Please, not that the best routes in our study are calculated using the Dijkstra's shortest path algorithm.

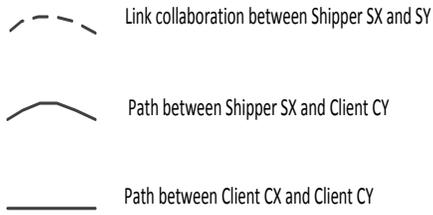

Link collaboration between Shipper SX and SY

Path between Shipper SX and Client CY

Path between Client CX and Client CY

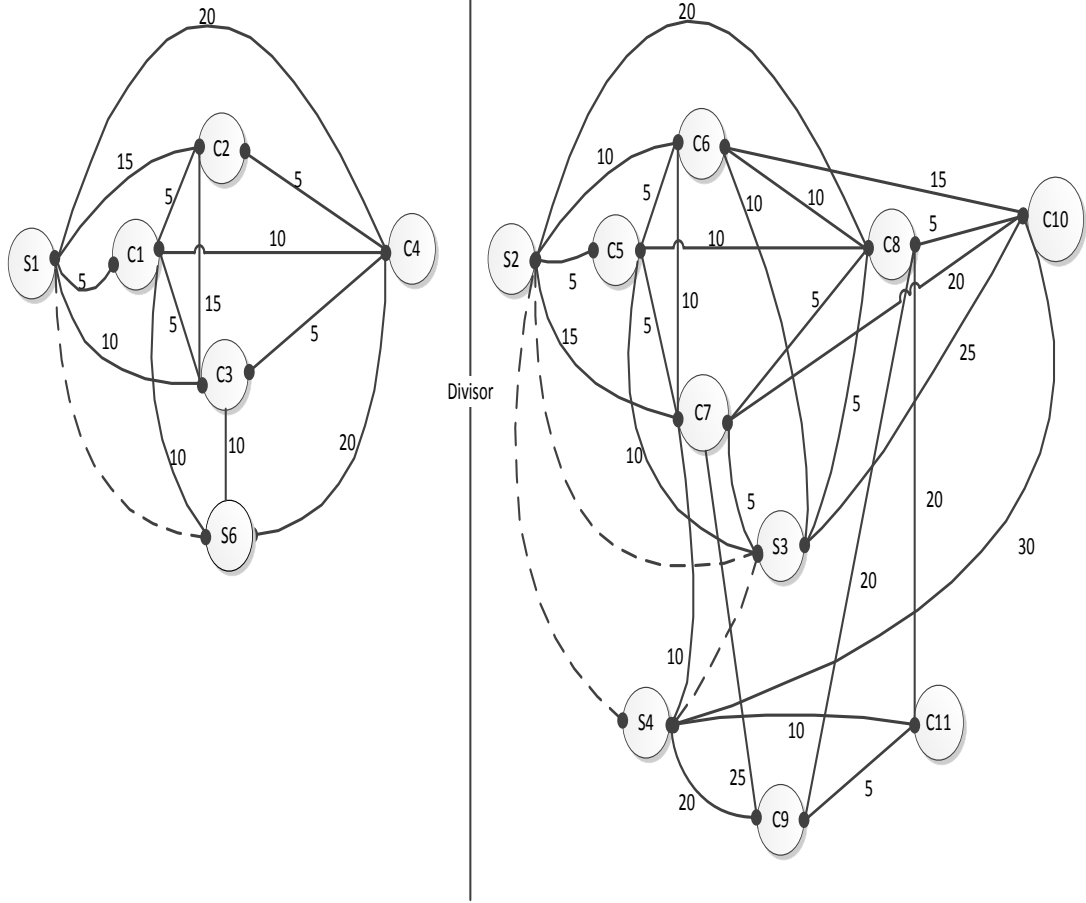

**Fig.13: Shippers and clients location –Network diagram**

## Table 13: Schedules and possible routes

| Collaboration type | Possible combinations between shippers | Travel time | Clients | Route used (with shortest path) | Number of trips | Total triplength | Kinds of vehicles used(counts) |
|---|---|---|---|---|---|---|---|
| FC | S1-S6 | ≤ 9-1pm | C2 | (S1-S6)-C2(15) | 2 | 30 | T1, T2 |
| | | | C1,C3,C4 | (S1-S6)-C1-C3-C4(15) | | | |
| PC | S2-S3 | ≤ 9-10am | C5 | (S2-S3)-C6(10) | 2 | 25 | T4, T5 |
| | | | C6, C7, C8 | (S2-S3)-C5-C7-C8(15) | | | |
| | S2-S4 | ≤ 9-10am | C7, C8 | (S2-S4)-C7-C8(15) | 1 | 15 | T4 or T7or T8 |
| | S2-S3-S4 | ≤ 9-10am | C7, C8 | (S2-S3-S4)-C7-C8(10) | 1 | 10 | T4 or T5 or T7or T8 |
| NC | S1 | ≤ 9am-12pm | C2 | S1-C2(15) | 2 | 30 | T1,T2 |
| | | | C1, C3, C4 | S1-C1-C3-C4(15) | | | |
| | S2 | ≤ 9am -11am | C6, C8 | S2-C6-C7(20) | 1 | 35 | T4 |
| | | | C5, C7 | S2-C5-C8(15) | | | |
| | S3 | ≤ 9am-10am | C6 | S3 –C6(10) | 1 | 45 | T5 |
| | | | C7,C8 | S3-C7-C8(10) | | | |
| | | | C10 | S3-C10(25) | | | |
| | S4 | ≤ 9am-10am | C7, C8 | S4-C7-C8(15) | 2 | 40 | T7,T8 |
| | | | C9, C11 | S4-C9-C11(25) | | | |
| | S6 | ≤ 9am-12pm | C1 | S6-C1(10) | 2 | 30 | T1,T2 |
| | | | C3, C4, C2 | S6-C3-C4-C2(20) | | | |

Note

FC: Full collaboration

PC: Partial Collaboration

NC: No collaboration

*Step 4: Environmental Impact Assessment*

In this step, we evaluate the impact of vehicle goods movement on the city environment in terms of pollution and emissions ($CO_2$, $NO_x$ etc.) respecting the number of vehicle trips and the total triplength that favor minimum vehicular emissions. Table 14 presents the results of these findings.

**Table 14: Vehicular emissions**

| Types of collaboration | Possible combinations between shippers | Route used (with shortest path) | Number of vehicles used | Total triplength | Kinds of vehicles used(counts) | Vehicular emissions |
|---|---|---|---|---|---|---|
| FC | S1-S6 | (S1-S6)-C2(15) | 2 | 30 | T1, T2 | E1 + E2 |
| | | (S1-S6)-C1-C3-C4(15) | | | | |
| PC | S2-S3 | (S2-S3)-C6(10) | 2 | 25 | T4, T5 | 2E1 |
| | | (S2-S3)-C5-C7-C8(15) | | | | |
| | S2-S4 | (S2-S4)-C7-C8(15) | 1 | 15 | T4 or T7or T8 | E1 or E2 |
| | S2-S3-S4 | (S2-S3-S4)-C7-C8(10) | 1 | 10 | T4 or T5 or T7or T8 | E1 or E2 |
| NC | S1 | S1-C2(15) | 2 | 30 | T1,T2 | E1 + E2 |
| | | S1-C1-C3-C4(15) | | | | |
| | S2 | S2-C6-C7(20) | 1 | 35 | T4 | 2E1 |
| | | S2-C5-C8(15) | | | | |
| | S3 | S3 –C6(10) | 1 | 45 | T5 | 3E1 |
| | | S3-C7-C8(10) | | | | |
| | | S3-C10(25) | | | | |
| | S4 | S4-C7-C8(15) | 2 | 40 | T7,T8 | E1+E2 |
| | | S4-C9-C11(25) | | | | |
| | S6 | S6-C1(10) | 2 | 30 | T1, T2 | E1 + E2 |
| | | S6-C3-C4-C2(20) | | | | |

Note

NC: No collaboration

PC: Partial collaboration

FC: Full collaboration

In numerical terms, we can say that

Vehicular emission of FC + Vehicular emission of PC < Vehicular emission of NC

: (E1 + E2) +2E1 + (E1 or E2) + (E1 or E2) < (E1 + E2) + 2E1 +3E1 + (E1+E2) +

(E1 + E2)

Hence, we infer that vehicular emissions within a city can be reduced through partial and/or full collaboration between shippers meeting the demands of clients with goods vehicle delivery.

## 4.2. Verification of model results

In the verification for the collaboration square model a new test sample having six tests cases was used; while, for the operational level model four shippers each serving four clients were considered. The purpose of the verification is to access the credibility of the applications of these models.

### 4.2.1 Verification for Macro-level (Collaboration square model)

Test sample 2 is presented as follows with six test cases. The selection made respecting the highest weighted collaboration intents for SN and CC are described as follow:

1. For SS, select test case 4 with collaboration matrix SS (5, 95, 30) and weighted collaboration intents of 130

2. For CC, select test case 5 with  collaboration matrix CC (30, 10, 120) and weighted collaboration intents of 160

## Test Sample 2

**Case 1:**

| | B | | | C | | | Collaboration matrix | | | Weight (%) |
|---|---|---|---|---|---|---|---|---|---|---|
| | S (+) | E (+) | En (+) | S (-) | E (-) | En (-) | | | | |
| B | 10 | 20 | 70 | 10 | 40 | 60 | SN(5 | 15 | -30) | SN(-10) |
| C | 15 | 55 | 30 | 20 | 30 | 50 | CC(-10 | -10 | 20) | CC(0) |

**Case 2:**

| | B | | | C | | | Collaboration matrix | | | Weight (%) |
|---|---|---|---|---|---|---|---|---|---|---|
| | S (+) | E (-) | En (+) | S (-) | E (+) | En (-) | | | | |
| B | 10 | 20 | 70 | 10 | 40 | 60 | SN(5 | -15 | -30) | SN(-40) |
| C | 15 | 55 | 30 | 20 | 30 | 50 | CC(-10 | 10 | 20) | CC(20) |

**Case 3:**

| | B | | | C | | | Collaboration matrix | | | Weight (%) |
|---|---|---|---|---|---|---|---|---|---|---|
| | S (+) | E (-) | En (+) | S (-) | E (+) | En (+) | | | | |
| B | 10 | 20 | 70 | 10 | 40 | 60 | SN(5 | -15 | 90) | SN(80) |
| C | 15 | 55 | 30 | 20 | 30 | 50 | CC(-10 | 10 | 120) | CC(120) |

**Case 4:**

| | B | | | C | | | Collaboration matrix | | | Weight (%) |
|---|---|---|---|---|---|---|---|---|---|---|
| | S (+) | E (+) | En (-) | S (-) | E (+) | En (+) | | | | |
| B | 10 | 20 | 70 | 10 | 40 | 60 | SN(5 | 95 | 30) | SN(130) |
| C | 15 | 55 | 30 | 20 | 30 | 50 | CC(-10 | 50 | -20) | CC(20) |

**Case 5:**

| | B | | | C | | | Collaboration matrix | | | Weight (%) |
|---|---|---|---|---|---|---|---|---|---|---|
| | S (+) | E (-) | En (+) | S (+) | E (+) | En (+) | | | | |
| B | 10 | 20 | 70 | 10 | 40 | 60 | SN(40 | -30 | 110) | SN(120) |
| C | 15 | 55 | 30 | 20 | 30 | 50 | CC(30 | 10 | 120) | CC(160) |

**Case 6:**

| | B | | | C | | | Collaboration matrix | | | Weight (%) |
|---|---|---|---|---|---|---|---|---|---|---|
| | S (-) | E (-) | En (-) | S (+) | E (+) | En (+) | | | | |
| B | 10 | 20 | 70 | 10 | 40 | 60 | SN(-5 | -30 | 30) | SN(-5) |
| C | 15 | 55 | 30 | 20 | 30 | 50 | CC(10 | -10 | -20) | CC(-20) |

The following Figure 14 describes the line plots for SN and CC weighted collaboration intents for the six test cases in Test sample 3.

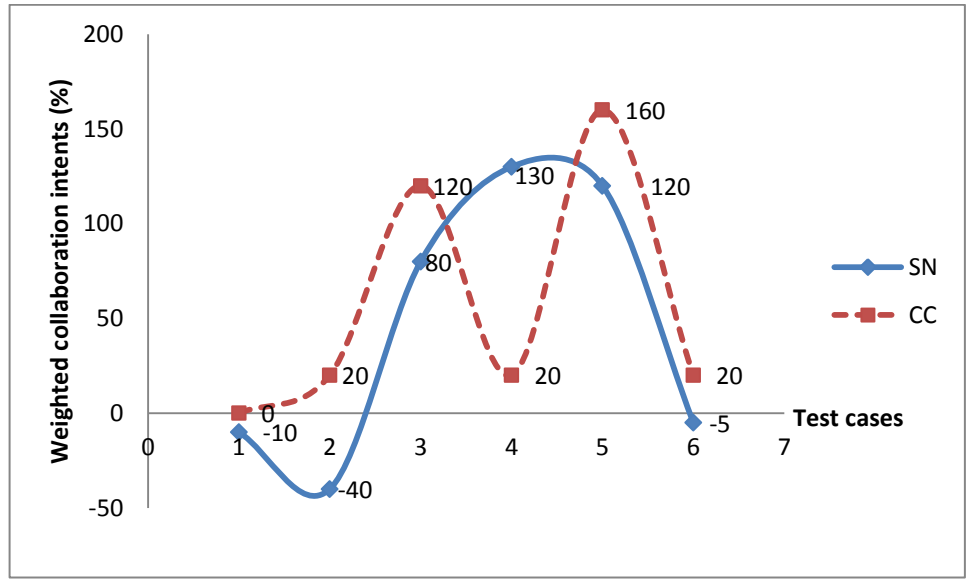

**Fig. 14: Line plots for SN and CC weighted collaborative intents versus test cases**

The verification for the collaboration square model tells us that:

1.  Despite the modification of data, we can rely on the collaboration square model in finding the optimum weighted collaboration intents for SN and CC.

2.  It does not necessarily imply that the maximum weighted collaborative intents could be found within a wider scope.

3.  The benchmark for the minimum number of test cases can be five for finding the local maximum for the weighted collaboration intents.

### 4.2.2 Verification for Micro-level (Operational level model)

In the verification for micro-level (operational level model), consideration is given only to CC. Table 15 describes the modified data for clients demand orders for (six) shippers

S1, S2, S3, S4, S5 and S6. Tables 15.1 to 15.6 contain modified data for trucks available for the shippers: S1, S2, S3, S4, S5 and S6.

The following assumptions have been made for the modeller:

1. The clients served by each of the shippers remain the same.

2. The delivery time windows remain constant.

3. The emission standards (ref. Table 10) remain valid.

**Table 15: Clients demand orders for shippers S1, S2, S3, S4, S5, and S6**

| Shippers | Clients | No of packets to be delivered(N) | Size of the Packet(S) | Time windows | Quantity to be delivered= (N*S) |
|----------|---------|----------|----------|----------|----------|
| S1 | C1 | 10 | 20 | 9am-10am | 200 |
|    | C2 | 50 | 2 | 10am-11am | 100 |
|    | C3 | 10 | 10 | 11am-12pm | 100 |
|    | C4 | 5 | 12 | 12pm-1pm | 60 |
| S2 | C5 | 6 | 20 | 9am-10am | 120 |
|    | C6 | 5 | 6 | 9am-10am | 30 |
|    | C7 | 20 | 10 | 9am-10am | 200 |
|    | C8 | 5 | 20 | 9am-12pm | 100 |
| S3 | C6 | 10 | 5 | 9am-10am | 50 |
|    | C7 | 10 | 6 | 9am-10am | 60 |
|    | C8 | 25 | 3 | 9am-12pm | 75 |
|    | C10 | 10 | 20 | 9am-10am | 200 |
| S4 | C7 | 12 | 5 | 9am-10am | 60 |
|    | C8 | 8 | 10 | 9am-12pm | 80 |
|    | C9 | 1 | 40 | 9am-10am | 40 |
|    | C11 | 2 | 30 | 9am-10am | 60 |
| S5 | C12 | 10 | 5 | 9am-10am | 50 |
|    | C13 | 20 | 4 | 9am-11am | 80 |
|    | C14 | 5 | 10 | 9am-11am | 50 |
|    | C15 | 10 | 20 | 9am-11am | 200 |
| S6 | C1 | 20 | 5 | 9am-10am | 100 |
|    | C2 | 10 | 2 | 10am-11am | 20 |
|    | C3 | 10 | 10 | 11am-12pm | 100 |
|    | C4 | 10 | 6 | 12pm-1pm | 60 |

**Table 15.1: Trucks available with S1**

| Trucks | Gains($) | Load capacity (kg) | Vehicle size(tons) | Emission Factor |
|--------|----------|--------------------|--------------------|-----------------|
| T1 | 500 | 200 | 160 | 1.5E1 |
| T2 | 100 | 50 | 160 | 1.5E2 |

**Table 15.2: Trucks available with S2**

| Trucks | Gains($) | Load capacity (kg) | Vehicle size(tons) | Emission Factor |
|--------|----------|--------------------|--------------------|-----------------|
| T3 | 80 | 100 | 300 | 1.2E2 |
| T4 | 90 | 50 | 200 | 1.2E1 |

**Table 15.3 Trucks available with S3**

| Trucks | Gains($) | Load capacity (kg) | Vehicle size(tons) | Emission Factor |
|--------|----------|--------------------|--------------------|-----------------|
| T5 | 1200 | 100 | 200 | 2E1 |
| T6 | 300 | 200 | 300 | 2E2 |

**Table 15.4 Trucks available with S4**

| Trucks | Gains($) | Load capacity (kg) | Vehicle size(tons) | Emission Factor |
|--------|----------|--------------------|--------------------|-----------------|
| T7 | 500 | 100 | 150 | 1.3E1 |
| T8 | 650 | 100 | 180 | 1.5E2 |

**Table 15.5: Trucks available with S5**

| Trucks | Gains($) | Load capacity (kg) | Vehicle size(tons) | Emission Factor |
|--------|----------|--------------------|--------------------|-----------------|
| T9 | 100 | 80 | 400 | 1.1E1 |
| T10 | 300 | 200 | 300 | 2E2 |

**Table 15.6: Trucks available with S6**

| Trucks | Gains($) | Load capacity (kg) | Vehicle size(tons) | Emission Factor |
|--------|----------|--------------------|--------------------|-----------------|
| T1 | 600 | 200 | 160 | 1.5E1 |
| T2 | 600 | 50 | 160 | 1.5E2 |

***Step 1: SCS Initial input filtering (Vehicle usage compliance model)***

In the Test sample, select test case 5 having collaboration matrix CC (30, 10, 120) with

weighted collaboration intents of 160. The constraints that can be set are:

C1: City restriction on vehicle size $\leq$ 600tons $\times$ 30%

$\leq$ 180tons

C2: Gains on distribution (for plausible taxation purpose) $\geq$ \$5000 $\times$ 10%

$\geq$ \$500

C3: Environment restriction on emission factor $\geq$ E1 $\times$ 120% *OR* E2 $\times$ 120%

The following Table 16 describes the input filtering for the shippers vehicles and the inference made based for each of these constraints.

### Table 16: Input filtering for Shippers' vehicles

| Shippers | Vehicles/Trucks | Constraints | | | Accept/Reject trucks | Constraints inference |
|---|---|---|---|---|---|---|
| | | C1 | C2 | C3 | | |
| S1 | T1 | √ | √ | √ | Accept | Fully satisfied |
| | T2 | √ | × | √ | Reject | Partially satisfied |
| S2 | T3 | × | × | √ | Reject | Partially satisfied |
| | T4 | × | × | √ | Reject | Partially satisfied |
| S3 | T5 | √ | × | √ | Reject | Partially satisfied |
| | T6 | × | × | √ | Reject | Partially satisfied |
| S4 | T7 | √ | √ | √ | Accept | Fully satisfied |
| | T8 | √ | √ | √ | Accept | Fully satisfied |
| S5 | T9 | × | × | × | Reject | Partially satisfied |
| | T10 | × | × | √ | Reject | Partially satisfied |
| S6 | T1 | √ | √ | √ | Accept | Fully satisfied |
| | T2 | √ | √ | √ | Accept | Fully satisfied |

√ : Satisfied
× : Not satisfied

From Table 16, the verification showed that only trucks T1 for S1; T7, T8 for S4; and T1, T2 for S6 meets the requirements of the constraints for vehicle usage in the city.

***Step 2: Goods to vehicle assignment***

Table 17.1 to 17.3 describes the goods assigned to trucks accepted for use by the shippers respecting the load capacity of the trucks.

**Table 17.1: Possible allocations of packets to truck T1 by S1**

| Load case number | Trucks with their loads | | | | Total number of vehicle trips |
|---|---|---|---|---|---|
| Case 1.1 | T1(100) | T1(200) | T1(100 + 60) | | 3 |
| Case 1.2 | T1(100) | T1(100 + 60) | T1(200) | | 3 |
| Case 1.3 | T1(100 + 60) | T1(200) | T1(100) | | 3 |
| Case 1.4 | T1(200) | T1(100 + 100) | T1(60) | | 3 |
| Case 1.5 | T1(100 + 100) | T1(60) | T1(200) | | 3 |
| Case1.6 | T1(200) | T1(60) | T1(100 + 100) | | 3 |
| Case 1.7 | T1(60) | T1(100) | T1(100) | T(200) | 4 |
| Case 1.8 | T1(200) | T1(100) | T1(100) | T(60) | 4 |
| Case 1.9 | T1(100) | T1(60) | T1(100) | T(100) | 4 |

**Table 17.2: Possible allocations of packets to trucks T7, T8 by S4**

| Load case number | Trucks with their loads | | | | Total number of vehicle trips |
|---|---|---|---|---|---|
| Case 2.1 | T7(60 + 40) | T7(80) | T8(60) | | 3 |
| Case 2.2 | T7(60 + 40) | T8(60) | T7(80) | | 3 |
| Case 2.3 | T7(80) | T8(60 + 40) | T8(60) | | 3 |
| Case 2.4 | T8(60 + 40) | T7(80) | T8(60) | | 3 |
| Case 2.5 | T8(60) | T8(80) | T7(60) | T7(40) | 4 |
| Case 2.6 | T8(60) | T7(60) | T7(40) | T8(80) | 4 |
| Case 2.7 | T7(60) | T8(60) | T8(80)) | T7(40) | 4 |
| Case 2.8 | T7(40) | T8(80) | T7(40) | T7(60) | 4 |
| Case 2.9 | T8(60) | T8(60) | T8(40) | T7(60) | 4 |

**Table 17.3: Possible allocations of packets to trucks T1, T2 by S6**

| Load case number | Trucks with their loads | | | | Total number of vehicle trips |
|---|---|---|---|---|---|
| Case 2.1 | T1(60 + 100) | T2(100) | T2(20 ) | | 3 |
| Case 2.2 | T1(20 + 100) | T2(60) | T1(100) | | 3 |
| Case 2.3 | T2(100 + 20) | T1(100+60) | | | 2 |
| Case 2.4 | T1(100+60) | T2(100 + 20) | | | 2 |
| Case 2.5 | T1(60) | T2(100) | T1(20) | T2(100) | 3 |
| Case 2.6 | T1(100) | T1(60) | T2(100) | T2(40) | 4 |
| Case 2.7 | T1(60) | T2(100) | T2(20)) | T1(100) | 4 |

For the goods to vehicle assignment, the verification revealed that minimum number of vehicle trips is achievable based on the allocation of packets to trucks. It can be said Truck T1 can make a minimum, three trips for S1; T7 can make minimum, three trips and T8 for S4 and T1, T2 can make minimum two trips for S6.

### *Step 3: Vehicle routing and scheduling*

The network path for the routing and scheduling is shown in Figure 15. The following Table 18 describes the routing and scheduling of the vehicles according to the TSP.

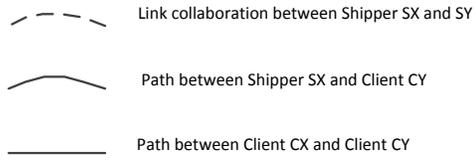

Link collaboration between Shipper SX and SY

Path between Shipper SX and Client CY

Path between Client CX and Client CY

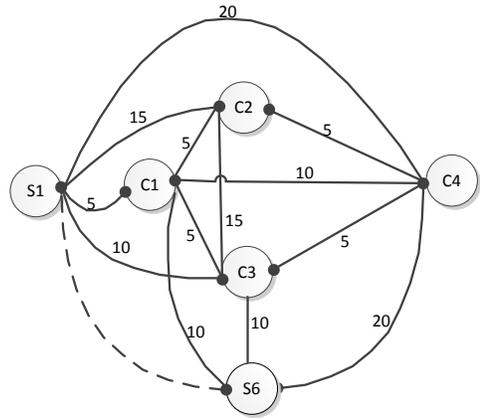

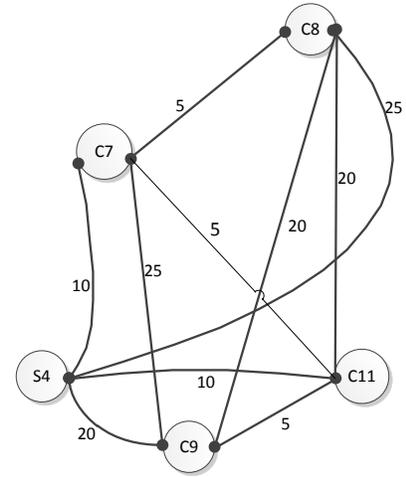

Divisor

**Fig. 15: Shippers and clients location −Network diagram**

**Table 18: Schedule and possible route**

| Collaboration type | Possible combinations between shippers | Travel time | Clients | Route used (with shortest path) | Number of trips | Total triplength | Kinds of vehicles used(counts) |
|---|---|---|---|---|---|---|---|
| FC | S1-S6 | ≤ 9-1pm | C2 | (S1-S6)-C2(15) | 2 | 30 | T1, T2 |
| | | | C1,C3,C4 | (S1-S6)-C1-C3-C4(15) | | | |
| NC | S1 | ≤ 9am-12pm | C1 | S1-C1(5) | 3 | 35 | T1 |
| | | | C2 | S1-C2(15) | | | |
| | | | C3, C4 | S1-C3-C4(15) | | | |
| | S4 | ≤ 9am-10am | C7,C8 | S4-C7-C8(15) | 3 | 45 | T7,T8 |
| | | | C9 | S4-C9(20) | | | |
| | | | C11 | S4 -C11(10) | | | |
| | S6 | ≤ 9am-12pm | C1,C2 | S6-C1-C2(15) | 2 | 30 | T1,T2 |
| | | | C3,C4 | S6-C3-C4(15) | | | |

Note
FC: Full collaboration
PC: Partial Collaboration
NC: No collaboration

As can be seen from Table 18, it can be said that the with full- collaboration (FC), the total trip length can be less than that with no-collaboration (NC). In numerical terms, this can be describes as:

Total trip lengths with FC < Total trip lengths with NC

: FC (S1-S6) < NC (S1) +NC (S6)

30 < (35 +30)

For shipper S4, it can be said that the minimum number of trips to be made by the shipper with goods delivery to clients C7, C8, C9 and C11 is unlikely to be reduced respecting vehicle routing and scheduling in accordance to TSP.

*Step 4: Environmental impact assessment*

The following Table 19, describe the vehicular emissions for the shippers' trucks under FC and NC.

**Table 19: Vehicular emissions**

| Types of collaboration | Possible combinations between shippers | Route used (with shortest path) | Number of vehicles used | Total triplength | Kinds of vehicles used(counts) | Vehicular emissions |
|---|---|---|---|---|---|---|
| FC | S1-S6 | (S1-S6)-C2(15) | 2 | 30 | T1, T2 | 1.5E1 + 1.5E2 |
| | | (S1-S6)-C1-C3-C4(15) | | | | |
| NC | S1 | S1-C1(5) | 3 | 35 | T1,T2 | 1.5E1 + 1.5E2 |
| | | S1-C2(15) | | | | |
| | | S2-C3-C4(15) | | | | |
| | S4 | S4-C7-C8(15) | 3 | 45 | T7,T8 | 1.3E1+ 1.8E2 |
| | | S4-C9 (20) | | | | |
| | S6 | S6-C11(10) | 2 | 30 | T1, T2 | 1.5E1 + 1.5E2 |
| | | S6-C3-C4-C2(20) | | | | |

Note

NC: No collaboration

PC: Partial collaboration

FC: Full collaboration

In numerical terms, we can say that for shippers S1, S6

Vehicular emission of trucks with FC < Vehicular emission of trucks with NC

: (1.5E1 + 1.5E2) +) < (1.5E1 + 1.5E2) + (1.5E1 + 1.5E2)

Hence, it can be inferred that with the modeller, vehicular emissions within a city can be reduced through full collaboration between shippers as it was in the case of the model.

For shippers S4, the verification failed to provide a ground for comparison since the shipper has a one-to-many clients relationships. This implies that the operational level model would be more effective in a many shippers-to-many clients' relationships.

### 4.2.3. Conclusion on verifications of models

The results obtained from the verifications have shown that the result of the verification compares reasonably well as that of the actual models –for the collaboration square model and operational level model –confirming their credibility and reasonability.

## 4.3. Validation of model results

### 4.3.1. Collaboration square model

Presently, there are no real systems or existing model for the validation of the collaboration square model.

### 4.3.2. Operational level model

The validation of the operational level model can be done using an existing model known as the Awasthi and Proth' CL decision making model (2006). In the approach of the CL decision making model and the operational model; it can be said that while the CL decision making model is best suited for one shipper-to-many clients' relationship, the operational level model performs well in a many shipper-to-many clients' relationship.

Also, the operational level model affirms the necessity of partial and full collaboration for city operators in CL planning.

# Chapter 5:

# Conclusions and future work

## 5.1. Conclusions

In this research, we have critically examined the subject matter of collaboration planning of stakeholders for achieving sustainable CL operations as related to the collaboration questions that have often been of interest to academicians and practitioners on CL planning and supply chain. We provided solutions approach that can help stakeholders achieve this goal. Two categories of models are proposed to evaluate the collaboration strategies. At the macro level, we have the simplified collaboration square model and advance collaboration square model and at the micro level we have the operational level model. These collaboration decision making models, with their mathematical elaborations on business-to-business, business-to-customer, customer-to-business, and customer-to-customer provide roadmaps for evaluating the collaboration strategies of stakeholders for achieving sustainable city logistics operations attainable under non-chaotic situation and presumptions of human levity tendency.

By our numerical analysis, we reached a conclusion that full collaboration can be optimized when two or more shippers have equal number of clients that collaborate and partial collaboration can be optimized when two or more, or all shippers collaborate to serve a client or clients respecting the minimum number of vehicle trips and total trip length that favor minimum vehicular emissions. The summary of this research endeavor is presented in the following Figure 16.

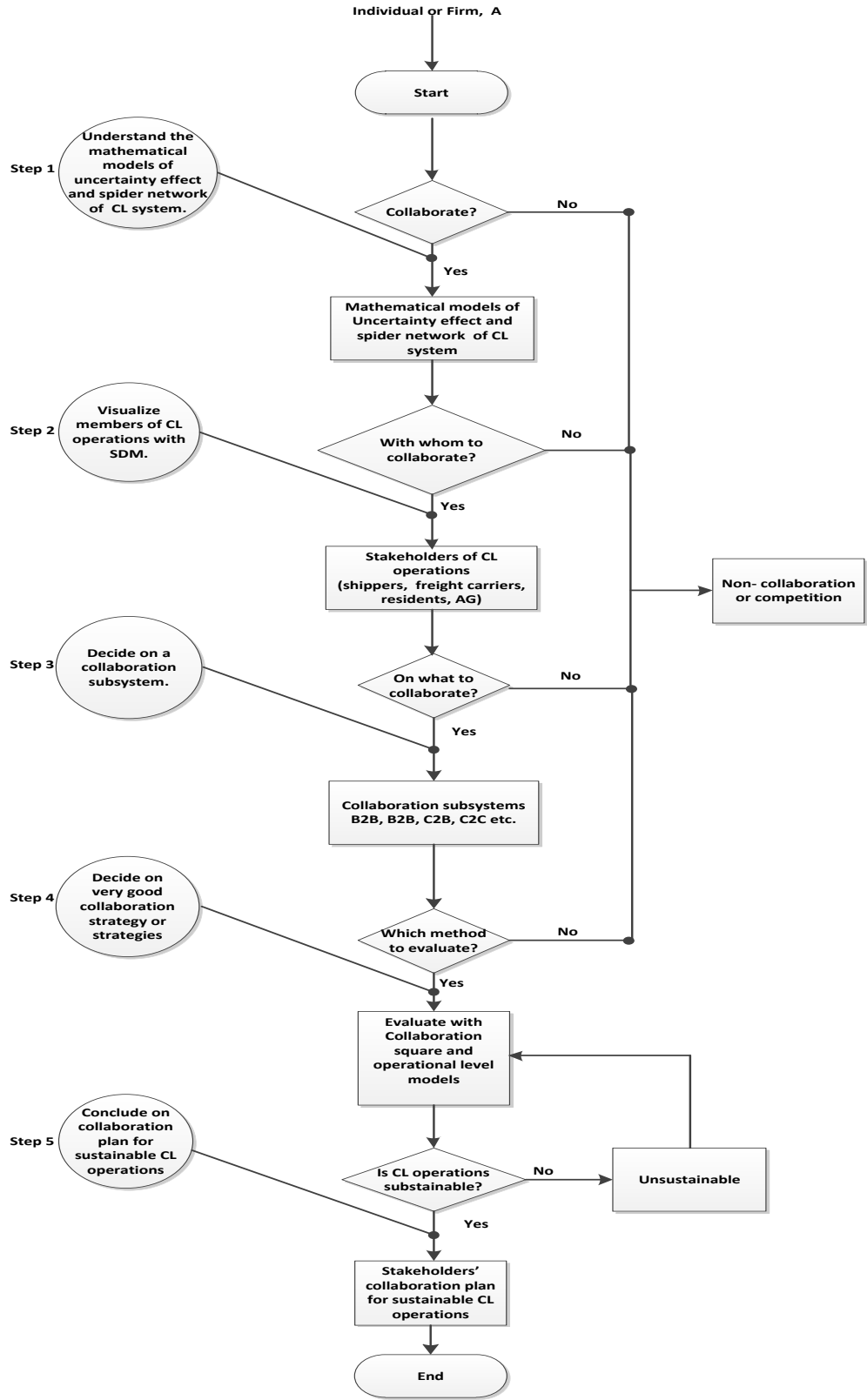

**Fig. 16: Flowchart of solutions approach for sustainable CL operations**

The results of SWOT analysis for the proposed work are described as follows:

**Strengths include:**

1. Providing a conceptualization of uncertainty effect and the dynamic complexity of CL.

2. Developing models for evaluating stakeholder collaboration strategies with focus on the basic and major collaboration subsystems comprising of B2B, B2C, C2B and C2C

3. Finding the application of the models at the macro and micro-level of CL systems.

4. Affirming the relevance of partial and/or full collaboration planning of stakeholders for achieving sustainable CL systems.

**Limitations include:**

1. The hypothetical values used to describe the negative and positive effectors of uncertainty effect are not realistic.

2. The collaboration square models is presently limited for explaining B2B, B2C, C2B and C2C subsystems in relation to social network and consolidation centers and may be untrue for other subsystems.

3. The test cases used in the Test samples for the numerical analysis at macro-level have been limited to five and six test cases respectively. However, better results could be achieved if the maximum number of test cases can be sixty-four, for finding the global maximum rather than the local maximum used in this case.

4. The application of our models at the operational level of CL systems is limited to only consolidation centers. This work does not cover the application of the models to social networks.

**Opportunities include:**

1. The application of the simplified collaboration square model at the macro-level for accessing the impacts of social-cultural characteristics, economy and environment provide a new theoretical bases for assessing the collaborative intents of stakeholders within the framework of B2B, B2C, C2B and B2B alliance towards the formation of consolidation centers and social networks.

2. The models can be useful for stakeholders aiming to reduce pollution in the city through partial and full collaboration.

3. The operational level models can help stakeholders develop a better framework for respecting legislations on vehicle usage in cities, assignment of goods to vehicles, vehicle routing and scheduling respecting delivery time windows, environmental impact assessment involving many shippers to clients' relationship.

**Threats:**

1. The hypothetical values assigned to the positive and negative effectors of uncertainty, the presumptions of human levity tendencies and the non-chaotic situations of the trio-conditionality used in this thesis are heuristics that have not been proven experimentally and should be used with caution in any human related research.

## 5.2. Future work

In future, we will advance our present work to other areas of applications in city logistics operations. Specifically, we will explore in more detail the collaboration square models using game theory with regards to decision making by stakeholders. Also, we shall develop software program to analyze larger number of test cases for our collaboration square models.

The scope of this research has been limited to B2B, B2C, C2B, and C2C subsystems that form the basic and major means of performing e-commerce or collaboration. For future work, exploring the usability of these models for others subsystems and supply chain is recommended.

# Glossary

Axiom:  A self-evident truth that requires no proof.

Admissible heuristics: It is also known as optimistic heuristics. A heuristics is admissible if the cost of reaching the goal is within estimate (Russell and Norvig (2002)).

Chaotic situation: In mathematics, it can be simply defined as randomness; and in physics, it can be defined as any state of confusion or disorder in the behavior of certain nonlinear dynamical systems.

Complexity: The quality or state of being complex or very difficult. For instance, NP-complete problems are regarded as polynomial mathematical problems with very high complexity.

Consolidation center (CC):  A center where business owners agree to merge their products to foster their commercial interest.

Deterministic: A valid outcome of preexisting sufficient causes.

Dynamic: Characterized by continuous change.

Effect: Something that is produced by cause; consequence; or result.

Flowchart: A type of diagram that represents an algorithm or process, showing the steps as boxes of various kinds.

Heuristic: An experimental method base on trial and error that serve as an aid to learning, problem-solving or discovery.

Human levity tendency (HLT): The behavioral way of assuming all is normal within a human set boundary despite contrary evidence pointing to events, elsewhere, outside the boundary.

Hypothetical value: A value assumed to exist as an immediate consequence of a hypothesis.

Logistics: Is the management of the flow of goods between the point of origin and the point of destination in order to meet the requirements of customers or corporations.

Social networking (SN): An online web platform or site for networking people around the world and which is fast becoming one of the most lucrative ecommerce site e.g. Facebook.

Stochastic: Characterized by chance or probability.

Test case: It is case contained in the test sample for study, verification and validation of a model.

Uncertainty effect: Inability to predict the nature of effect of a future state of an object or event.

# End Notes